\documentclass[prd,aps,showpacs,superscriptaddress,nofootinbib,notitlepage,floatfix]{revtex4-2}
\usepackage{epsfig,graphicx,amsfonts,amsmath,amssymb,amsthm,dsfont}

\DeclareMathAlphabet{\msfsl}{OT1}{cmtt}{m}{sl}
\RequirePackage{mathrsfs}
\usepackage{color}
\usepackage{epsf}
\usepackage{epstopdf}

\newcommand{\mO}{\mathcal{O}}
\newcommand{\mR}{\mathcal{R}}
\newcommand{\mE}{\mathcal{E}}
\newcommand{\bperp}{\perp}
\newcommand{\tr}{\mathrm{tr}}
\newcommand{\bW}{\mathbb{W}}
\newcommand{\bG}{\mathbb{G}}

\newcommand{\bU}{\mathbb{U}}
\newcommand{\bV}{\mathbb{V}}
\newcommand{\bone}{\mathds{1}}
\def\eq#1{{Eq.~(\ref{#1})}}
\newcommand{\Le}{\left(}
\newcommand{\Ra}{\right)}

\newcommand{\beq}{\begin{equation}}
	\newcommand{\eeq}{\end{equation}}
\newcommand{\beqar}{\begin{eqnarray}}
	\newcommand{\eeqar}{\end{eqnarray}}
\newcommand{\D}{\partial}

\newcommand{\mB}{\mathcal{B}}
\newcommand{\mA}{\mathcal{A}}

\newcommand{\mV}{\mathcal{V}}

\begin{document}
	\title{Ordered exponentials and the effective currents of high energy gravity}
	\author{S. Bondarenko}
	%\author{M. A. Zubkov}
	\affiliation{Ariel University, Ariel 4070000, Israel}

%\date{\today}
	
\begin{abstract}

 We derive a Lipatov type effective action for the Reggeized gravitons. In place of the order
by order construction of the effective currents from diffeomorphism invariance, we propose a current which is
a single ordered exponential $W_{ij}$ along the light cone, generated by the optical tidal
matrix integrated once. The nested ordering of it reproduces the known
perturbative answer with some new terms added. These new terms are
non-local along the light cone and have a structure which the diffeomorphism construction cannot produce, the form of the third order contribution to the current
is also calculated. The
effective action, the current to Reggeon vertex and the current to current interaction are obtained
in closed form, in three equivalent writings, one of them covariant. Generally defined $W_{ij}$ reduces
to four parameters of the ray bundle, its density, its shear, the orientation of the shear and a
rotation of the image. The effective current uses three of them, being the second light cone
derivative of a single scalar, it is not sensitive to the orientation. Comparing with the standard eikonal
gravitational Wilson line, which is the phase, $W_{ij}$ is the Hessian of the same eikonal: it
carries the shear and the focusing that the Abelian phase does not, and it is inverse to the
Jacobi propagator of the congruence at first order in the field.
A generalization of the proposed ordered operator is discussed as well as applications of the gravitational effective action.

\end{abstract}
\maketitle

\section{Introduction}

 Reggeization of the $t$-channel scattering amplitudes at high energy is a well known and thoroughly studied result in quantum field theory,
see \cite{Regge,BFKL,BFKL1,BFKL2,LipatovEff,LipatovEff1,LipatovEff2}. Generally, in different scattering processes the Reggeization manifests itself in the appearance of new degrees of freedom,
called Reggeons, whose propagator differs from the propagator of the regular particles.
In QCD, for example, the Reggeization appears as Reggeized gluons and quarks, with the theory reformulated in terms of them.
There a colorless bound state of two Reggeized gluons, the Pomeron, provides the leading contribution to the processes with vacuum quantum numbers in the $t$-channel. An account of the interactions between Pomerons or Reggeons, in turn, leads to the construction of Regge Field Theory (RFT) for the corresponding degrees of freedom, see
\cite{LipatovEff,LipatovEff1,LipatovEff2,Our,Our01}, which is a generalization of the Gribov Reggeon approach, \cite{Gribov} for the case of QCD. The same holds for gravity, the Reggeization of the amplitudes with $t$-channel graviton exchange between the scattered particles was studied as well,
see \cite{LipatovGrav,LipatovGrav01,LipatovGrav02,LipatovGrav03,LipatovGrav04,LipatovGrav05,LipatovGrav06,LipatovGrav1,LipatovDL}.

 An important issue behind the calculations is the formulation of a unitary approach to the calculation of the scattering amplitudes. In QCD the calculations of this type can be done in the framework of the initial BFKL (Balitsky, Fadin, Kuraev, Lipatov) approach, 
see \cite{BFKL,BFKL1,BFKL2,Fadin,Fadin1,Fadin2,Fadin3,Fadin4,Fadin5,Fadin6}, or in the framework of Lipatov's effective action, see
\cite{LipatovEff,LipatovEff1,LipatovEff2,LipatovGrav,LipatovGrav01,LipatovGrav02,LipatovGrav03,LipatovGrav04,LipatovGrav05,LipatovGrav06,LipatovGrav1,LipatovDL} for the QCD and gravity frameworks. The advantage of the latter formalism is that the calculations can then be formalized on the basis of powerful QFT methods, see 
\cite{Our,Our01,Our1,Our11,Our12}. Practically, the effective action describes a local interaction, ordered in rapidity space, of two truncated propagators, i.e. of
P-ordered Wilson lines or ordered exponentials, see \cite{OurZubkov1,Our11,OurZubkov2,Ivanov}, and then the Reggeization appears as a consequence of the form of these ordered non-local operators. The Reggeized particles appear in this construction as auxiliary ones, see the details in \cite{OurZubkov1} for the QCD case and in 
\cite{OurZubkov2} for the case of the Einstein-Cartan gravity set-up. Nevertheless, the notion of the Reggeized degrees of freedom is different in QCD and in Einstein-Hilbert gravity.

 The first problem we face then is that the Einstein-Hilbert action carries no internal
gauge algebra, unlike the Einstein-Cartan formulation, where the local Lorentz group supplies
one and the Reggeization appears in the same way as in QCD. The theory is of course a gauge
theory of diffeomorphisms, but the diffeomorphism generators do not sit inside the line in the
way the color ones do. Namely, so far, to the best of our knowledge, in the Einstein-Hilbert formulation a Wilson line ordered in rapidity space and depending on the graviton degrees of freedom has not been defined or discussed\footnote{Of course the regular Wilson line is present in the high energy gravity constructions, see 
for example \cite{Alessio} for resent BFKL based calculations of gravitational amplitudes where the regular Wilson line is introduced, see also further.}. The formulation of the Reggeization there was then based on a bottom-up perturbative calculation of the vertices
of the Reggeon-particle interactions needed for the calculation of the trajectory of the Reggeized graviton, see \cite{LipatovGrav,LipatovGrav01,MyGrav}. This task is technically very involved and it has in addition a problem with the verification of the results. The method rests on the diffeomorphism invariance of the proposed form of the interaction of the Reggeized gravitons with the effective currents, and it is therefore sensitive to the form of the currents introduced. The results of the calculations can of course be checked a posteriori, the trajectory of the Reggeized graviton is known and it may or may not be reproduced, but in general it is not clear whether the corresponding kernel misses some parts which do not affect that result but could affect the trajectories of some bound states of Reggeized gravitons.

 Secondly, this way of calculating the kernel of the equation for the Reggeized degrees of freedom is very complex and, because we talk about the ordering of non-local operators, it runs into difficulties in the definition of the complex expressions. In particular, it does not define the action of nested non-local operators on functions and functionals. Also, again, performing the calculations in this way we do not know whether we miss some parts of the expressions, simply because of the method applied. The closed form of the gravity Lipatov effective action then cannot be written, we simply do not know the form of the Lipatov effective currents we have to use there. What is clear anyway is that the effective currents we need cannot be the regular Wilson lines defined through the graviton components in the usual way. Such a Wilson line does not account for the ordering in rapidity space and then, simply by construction, cannot be used directly for the calculation of the Lipatov effective vertices in the effective action approach.
 
 Technically, the problem we try to resolve is the following. In QCD, the P-exponential is an infinite series of the matrices in color space, and the form of the matrices depends on the particular representation of the symmetry group. Because the group is non-Abelian the matrices do not commute, and this is the reason why the P-ordering appears in the QCD Wilson lines. For the Einstein-Hilbert action the situation is different. In the high energy limit we discuss, the metric components we have are
$h_{\pm\pm}$, and the Wilson line we obtain is simply an integral of the scalar $p^{\pm} p^{\pm} h_{\pm \pm}$ along the path, i.e. it is an Abelian expression which provides no ordering and requires no P ordering in general.
Thus, the non-commutativity we need is not in the phase but it exists in the transport of the transverse deviation vector across the line. The object ordered along the line is a $GL(2,\mathbb{R})$ matrix on the two dimensional screen, generated by the first light cone primitive of the tidal matrix taken with the opposite sign, 
$-\D_{+}^{-1}R_{i+j+}$. Its generators taken at different $x^{+}$ do not commute whenever the principal axes of the tidal matrix rotate along the line, see 
\eq{SecG2101} further in the text.
Therefore, we introduce this new operator in the theory, constructed in the P-ordered way similarly to what is done in QCD, which is expected to reproduce the known trajectory of the Reggeized graviton\footnote{The check we do not perform here, see the discussion in Section~\ref{SecJ}.} and to clarify the structure of the effective action and the related geometry. This is the task performed in the article.

  Constructing the requested ordered operators we obtained results closely related to geometry. Firstly, in this framework two ordered objects
appear. They are similar at first sight but in general they have to be kept apart. One is generated by minus the first light cone primitive of
the tidal matrix and it carries the effective current. The other is the Jacobi propagator of the null congruence, generated by the deformation matrix
which solves the Riccati equation with the curvature. The two are inverse to each other at the first order in $h$, transporting the deviation form
and the deviation vector correspondingly, and the inversion fails beyond it. Secondly, the
ordering on the two dimensional screen, which is the stage of the formalism and which is orthogonal to the high energy beam of the particles, is resolved in a closed form. The transport introduced is fixed there by four parameters: the density of the ray bundle, the accumulated shear, the orientation of the
shear and the rotation of the image. The usual effective current of the effective action reads three of them and does not see
the orientation. We mention as well that these four parameters are the standard ones of light ray optics and of
gravitational lensing. The density is the van Vleck and Morette determinant itself, the zeros of
the Jacobi determinant inverse to it are the caustics and they supply the Morse index of the eikonal
amplitude, so the translation of the introduced operator to that literature is direct, see \cite{PenMao,Shore}.

 Therefore, the paper is organized as follows. In the next Section~\ref{SecB} we very briefly recall the main notations of \cite{LipatovGrav,LipatovGrav01,MyGrav} concerning the perturbative construction of the gravitational effective currents on the basis of diffeomorphism invariance. In Section~\ref{SecC} we propose a closed form of the operators requested for the construction of the effective currents. The operators provide a P-ordered expansion and then clarify the form of the Reggeon-graviton interaction vertices at each perturbative order. The operators introduced allow different forms of the generator. There is a simplest LO form based on the $h_{\pm\pm}$ components of the metric only, and there is a full geometrical form constructed from the Riemann tensor; the relations between the generators are clarified in the Section. Section~\ref{SecD} is dedicated to the verification of the results. We check that the known results are reproduced, check the correctness of the previous second order answer for the current and calculate the third order interaction vertices. In this Section we use the LO form of the proposed generator. In Section~\ref{SecE} we then present closed forms of the interaction term of the effective action written through the Reggeon-graviton interactions, as well as the action written as a direct interaction of the introduced operators. In Section~\ref{SecF} we discuss the properties of the formalism using the geometrical form of the generator and generalize the results obtained before. Section~\ref{SecG} is dedicated in turn to the covariant generalization of the formalism. There we discuss the definition of
the proposed ordered exponential beyond the light cone coordinates and beyond the standard effective currents, on the basis of the covariant definition of the currents and of the irreducible decomposition of the generator in terms of the light ray optical invariants. In Section~\ref{SecH} we write the effective action in this covariant form, with and without Reggeons, and discuss the difference between the operators of the formalism and the regular gravitational Wilson lines based on the Abelian phase. After that, in Section~\ref{SecI} we investigate the calculation schemes using the simplest shock wave profile for the graviton $h_{\pm \pm}$ components. In Section~\ref{SecJ} we briefly present the results obtained and the conclusions. The Appendices added to the text collect some technical calculations; the references to the individual Appendices are given at the corresponding places in the text.

\section{Set-up for the Lipatov's effective action for Reggeized gravitons}\label{SecB}

 In this Section we briefly recall some results and notations based on the \cite{MyGrav} paper and requested for the consistency and understanding of the further calculations.
We discuss the t-channel high energy gravitational amplitudes, universal for the interactions of different types of particles at high energy,
and restrict the consideration only to the part of any general action which describes the interaction of regular gravitons with the Reggeized ones:
\beq\label{SecB1}
L_{\rm int}\,=\,\frac{1}{2\kappa}\,\D_i^{2}\mB_{\mu\nu}\,j^{\,\mu\nu}(h)\,=\,\frac{1}{2\kappa}\,\D_i^{2}\mB_{\mu\nu}\Le h^{\mu\nu}\,+\,j^{\,\mu\nu}_{1}(h)\Ra\,=\,
\frac{1}{2\kappa}\,\mB_{\mu\nu}\,\D_i^{2} j^{\,\mu\nu}(h)\,,
\eeq
see more details in \cite{MyGrav}. The $j^{\,\mu\nu}(h)$ is a Lipatov's effective current, an analog of the QCD effective currents, see 
\cite{LipatovEff,Our,Our01}.
Being the counterpart of QCD currents, the $j^{\,\mu\nu}(h)$ was not defined in the same way nevertheless. It was introduced in \cite{LipatovGrav1} firstly in the context of
the gravity effective action but was never defined in the closed form, unlike the QCD counterpart. Instead, the form of the current was defined by perturbative order by order calculations following from the diffeomorphism invariance of the proposed interaction term and written as
\beq\label{SecB2}
\D_{i}^{2}\mB_{\mu \nu}\,\frac{\delta j^{\,\mu \nu}}{\delta h^{\rho \sigma}}\,\delta h^{\rho \sigma}\,=\,
\D_{i}^{2}\mB_{\mu \nu}\,\frac{\delta j^{\,\mu \nu}}{\delta h_{\rho \sigma}}\,\Le \varepsilon_{\rho;\,\sigma}\,+\,\varepsilon_{\sigma;\,\rho}\Ra\,=\,0\,.
\eeq
It was shown in \cite{LipatovGrav1,MyGrav}, that the equation could be resolved in terms of the following auxiliary variables
\beqar\label{SecB3}
X_{+i}\,&=&\,\D_{+}^{-1}\Le \D_{+}h_{+i}\,-\,\tfrac12\,\D_i h_{++}\Ra\,;
\\
Y_{ij}\,&=&\,\D_{+}^{-1}\Le \D_i h_{+j}\,+\,\D_j h_{+i}\,-\,\D_{+}^{-1}\D_i\D_j h_{++}\Ra
\,=\,\D_{+}^{-1}\Le \D_i X_{+j}+\D_j X_{+i}\Ra\,;
\label{SecB4}
\\
Z_{ij}\,&=&\,\tfrac12\,\D_{+}^{-1}\Le \D_i h_{+j}\,-\,\D_j h_{+i}\Ra\,,
\label{SecB5}
\eeqar
which acquire a simple LO form if we preserve in it only leading $h_{\pm \pm}$ component of the graviton field, see discussion in \cite{MyGrav}.
Then the effective currents, calculated till $h^{2}$ order, which provide leading contributions to the high energy scattering process are
\beq\label{SecB6}
j^{\mp\mp}=j_{\pm \pm}\,=\,h_{\pm\pm}\,+\,X_{\pm i}X_{\pm i}\,
\eeq
with sum over the $i$ assumed in the expression. Next, we solve LO classical equations of motion for the $h_{\pm\pm}$ and insert back the solution into the
gravity part of the full action obtaining
\beq\label{SecB7}
L_{\rm eff}\,=\,L_{EH}(\eta,\mB)\,+\,\frac{1}{2\kappa}\,j^{\,\mu\nu}(\mB)\,\D_i^{2}\mB_{\mu\nu}\,
\,-\,\frac{1}{2\kappa}\,\mB_{\mu\nu}\D_{\bperp}^{2}\mB^{\mu\nu}\,,
\eeq
the last term in the expression is the induced Reggeon field term.
That is the Lipatov effective action which describes the dynamics of the Reggeized graviton fields, the interactions of this field with other possible degrees of freedom provide in turn an effective action of particles whose gravitational interactions are described by these gravitons.

 There are two main problematic points in the construction we have. First, it is the completeness of the expression. Unlike the QCD case, the current is built
perturbatively from below, 
so terms which are invariant by themselves can be missed, this was already mentioned in \cite{MyGrav}.
Secondly, it is the ordering of the operators appearing in the expressions. In the calculation of the effective currents, already in the second order with respect to $h$,
the expression is
\beq\label{SecB8}
\delta j^{\mp\mp}/\delta X_{\pm i}\,=\,2X_{\pm i}\,.
\eeq
We see that the product $X_{+i}X_{+i}$ contains $\D_{+}^{-1}$ twice, and the relative light cone ordering
of the two inverse derivatives is not determined by the perturbative calculations. In the general set-up, see the QCD counterpart in \cite{Our,Our01}, 
the Reggeization comes from that ordering so this point is important. In QCD such problems do not arise, there the current is the light cone derivative of a Wilson
line and both gaps close at once including additionally a covariance of the currents in the Lagrangian. Therefore, the aim 
of the further calculations is the construction of the gravitational counterpart of the QCD effective current; i.e. we want one ordered exponential whose expansion
generates a closed form of the $j_{\pm\pm}$ with the ordering built in.

 Here we also clarify the notations we use throughout the paper. The signature used is $(+,-,-,-)$, the flat metric in light cone coordinates has
the following form
\beq\label{SecB9}
\eta_{+-}\,=\,\eta^{+-}\,=\,1\,,\qquad \,\eta_{ij}\,=\,-\delta_{ij}\,,\qquad i,j=1,2 .
\eeq
The transverse Laplacian is defined in the Euclidean sense
\beq\label{SecB10}
\D_{\bperp}^{2}\,\equiv\,\D_i\D_i=-\D_i\D^{\,i},
\eeq
which is the $\D_i^2$ used throughout the paper. Raising the light cone indices gives $h^{--}\,=\,h_{++}$. Also,
the transverse indices $i,j,k,l$ are Euclidean, always written downstairs and always summed with
$\delta_{ij}$, i.e. repeated indices are summed even when both are lower.
The gravity action is of Einstein-Hilbert
\beq\label{SecB1001}
g_{\mu\nu}\,=\,\eta_{\mu\nu}\,+\,h_{\mu\nu}\,,\qquad
L_{EH}\,=\,\frac{1}{\kappa}\,\sqrt{-g}\,R\,,\qquad
\kappa\,=\,16\pi G\,,
\eeq
so that the Einstein equations read $R_{\mu\nu}-\tfrac12 g_{\mu\nu}R\,=\,\tfrac{\kappa}{2}\,T_{\mu\nu}$ and,
in the Regge kinematics with $\eta_{\pm\pm}=0$, $R_{\pm\pm}\,=\,\tfrac{\kappa}{2}\,T_{\pm\pm}$ and
$\D_{\bperp}^{2}h_{\pm\pm}\,=\,\kappa\,T_{\pm\pm}$.
The linearized Riemann tensor \eq{SecF1} carries no $\kappa$ with this normalization, and
$\kappa$ enters the expressions only through $L_{EH}$ and through the source.
For the Reggeon sector the usual kinematic constraints are assumed:
\beq\label{SecB11}
\D_{\pm}\mB_{\mp\mp}\,=\,0 \,,\qquad
\D_{\mp}h^{\mp\mp}=\D_{\mp}h_{\pm\pm}\,=\,0\,;
\eeq
additional notations are also defined throughout the text where it is required.

\section{General notations: the operators $W_{ij}$ and $\mO$}\label{SecC}

 In our discussion we will also go from the bottom up. Namely, we have the answer for the effective currents calculated at second order precision, so
we could construct an ordered operator which will reproduce, at least partially, the answer. Further, investigating the construction, we can examine the proposed form of the operator
and its properties, as we will see we have not much freedom in this task. Now, first of all, let us introduce the notations we use in the calculations.
The objects we introduce below are $2\times2$ matrices in the transverse indices:
\beq\label{SecC1}
\bW\,=\,\{W_{ij}\}\,,\qquad (\bW_1\bW_2)_{ij}\,=\,W_{1\,ik}W_{2\,kj}\,,\qquad
\bone\,=\,\{\delta_{ij}\}\,,\qquad \tr\,\bW\,=\,W_{ii}\,;
\eeq
so the boldface notation denotes the
matrix and plain symbols with two indices denote its components.
Now, let $\bG\,=\,\{G_{ij}(x^{+},x_{\bperp})\}$ be a symmetric generator matrix, its particular
form is specified below, and let
\beq\label{SecC2}
\hat{G}_{ij}(x^{+},x_{\bperp})\,=\,\D_{+}^{-1}G_{ij}\,=\,
\int^{x^{+}}\!\!dx_{1}^{+}\;G_{ij}(x_{1}^{+},x_{\bperp})
\eeq
be its first light cone primitive, $\D_{+}\hat{G}_{ij}\,=\,G_{ij}$. The operator we work with
is the ordered exponential of $\hat{G}$,
\beq\label{SecC3}
W_{ij}(z^{+},x_{\bperp})\,=\,\Big[P\exp\int^{z^{+}}\!\!dx^{+}\,\hat{\bG}(x^{+},x_{\bperp})\Big]_{ij}\,,
\eeq
with $P$ ordering the integration variable, which is equivalent to the first order equation
\beq\label{SecC4}
\D_{+}W_{ij}\,=\,\hat{G}_{ik}\,W_{kj}\,,\qquad
W_{ij}\big|_{x^{+}\to\,-\,\infty}\,=\,\delta_{ij}\,.
\eeq
Differentiating \eq{SecC4} once more we obtain its second order form
\beq\label{SecC5}
\D_{+}^{2}W_{ij}\,=\,\Le G_{ik}\,+\,\hat{G}_{il}\hat{G}_{lk}\Ra W_{kj}\,,
\eeq
the second term in the bracket being absent if the ordering in \eq{SecC3} is dropped.
Equations \eq{SecC3} and \eq{SecC4} are formally the same as used in \cite{Our,Our01} for
the regular path-ordered Wilson line of the QCD effective action, with $\hat{G}_{ij}$ in place
of the gauge field.
The next operator we introduce is the one that reproduces the effective currents appearing in \eq{SecB1}. Following the QCD analogy, see \cite{Our,Our01}, we define
the new operator as
\beq\label{SecC6}
\mO_{++}\,=\,c\;\tr\big[\D_{+}^{2}\bW\big]\,=\,
c\,\D_{+}\tr\Le \hat{\bG}\,\bW\Ra\,=\,
c\,\Le G_{ik}\,+\,\hat{G}_{il}\hat{G}_{lk}\Ra W_{ki}\,,
\eeq
with the Euclidean trace assumed and $c$ a constant we need to reconstruct on the basis of the known perturbative answer.
We can solve \eq{SecC4} iteratively,
\beq\label{SecC7}
W_{ij}\,=\,\delta_{ij}\,+\,\Le \D_{+}^{-1}\hat{G}\Ra_{ij}\,+\,
\Big[\D_{+}^{-1}\Le \hat{G}\,\D_{+}^{-1}\hat{G}\Ra\Big]_{ij}\,+\,
\Big[\D_{+}^{-1}\Le \hat{G}\,\D_{+}^{-1}\Le \hat{G}\,\D_{+}^{-1}\hat{G}\Ra\Ra\Big]_{ij}\,+\,\dots\,,
\eeq
The integrations are nested, each $\D_{+}^{-1}$ acting on everything to its right, and this
nested light cone ordering is what distinguishes \eq{SecC7} from a product of independent
integrals. Inserting it into \eq{SecC6} we obtain the operator order by order in the generator.
Up to the third order we have
\beq
\mO_{++}=c\,G_{ii}+
c\,\Le G_{ik}\Le \D_{+}^{-1}\hat{G}\Ra_{ki}\,+\,\hat{G}_{ik}\hat{G}_{ki}\Ra+
%\nonumber
c\,\Le G_{ik}\Big[\D_{+}^{-1}\Le \hat{G}\,\D_{+}^{-1}\hat{G}\Ra\Big]_{ki}+
\hat{G}_{ik}\hat{G}_{kl}\Le \D_{+}^{-1}\hat{G}\Ra_{li}\Ra+\dots\,,
\label{SecC8}
\eeq
here the symmetric definition of the $\D_{+}^{-1}$ is assumed
$\D_{+}^{-1}f(x^{+})\,=\,\tfrac12\int dx_{1}^{+}\,\epsilon(x^{+}-x_{1}^{+})\,f(x_{1}^{+})$,
adopted in the QCD effective action, see \cite{Our}.

 The possible forms of the generator appearing in the \eq{SecC2} expression could be fixed as follows. First of all,
considering the simplest effective current based only on $h_{\pm \pm}$
graviton component, the task is trivial. In this case the generator is fixed by\footnote{For the sake of simplicity, further we will present the
expressions of only the $j_{++}$ component of the current, the $j_{--}$ component could be obtained then simply by changing the corresponding notations in the expressions. }
\beq\label{SecC9}
K_{ij}\,\equiv\, \D_i\D_j h_{++}\,,
\eeq
expression of \eq{SecB4}. So, denote
\beq\label{SecC10}
G^{(K)}_{ij}\,=\,\pm\,\tfrac12\,K_{ij}\,.
\eeq
Two signs are possible here and the first order term of \eq{SecC8} does not distinguish them but
the second order term does. As we will check further,
only the upper sign reproduces the $X_{+i}X_{+i}$ term of the perturbative current \eq{SecB6} and we take it throughout.
Rewriting the exponent of \eq{SecC3}, $\mE_{ij}\,\equiv\,\D_{+}^{-2}G_{ij}\,=\,\D_{+}^{-1}\hat{G}_{ij}$,
in terms of $K_{ij}$ we obtain
\beq\label{SecC11}
\mE^{(K)}_{ij}\,=\,\tfrac12\,\D_{+}^{-2}\D_i\D_j h_{++}\,,\qquad
W^{(K)}_{ij}\,=\,\Big[P\exp\Big(\tfrac12\D_{+}^{-2}\D_k\D_l h_{++}\Big)\Big]_{ij}\,.
\eeq
This is the LO form of the effective current, it contains $h_{++}$ only.

 Next define the same in terms of $Y_{ij}$ from \eq{SecB4}, in comparison to the previous case it contains sub-leading corrections as well. We have
\beq\label{SecC12}
\mE^{(Y)}_{ij}\,=\,-\,\tfrac12\,Y_{ij}\,,\qquad
G^{(Y)}_{ij}\,=\,-\,\tfrac12\,\D_{+}^{2}Y_{ij}\,=\,-\,\tfrac12\D_{+}\Le \D_i h_{+j}\,+\,\D_j h_{+i}\Ra\,+\,\tfrac12\D_i\D_j h_{++}\,,
\eeq
with
\beq\label{SecC13}
W^{(Y)}_{ij}\,=\,\Big[P\exp\Big(-\tfrac12\,Y_{kl}\Big)\Big]_{ij}\,,\qquad \mE^{(Y)}_{ij}\Big|_{h_{+i}\,=\,0}\,=\,\mE^{(K)}_{ij}\,.
\eeq
This expression reproduces the $h^{2}$ form of the current obtained in \cite{LipatovGrav,MyGrav}, plus additional terms as we will see further, written in the given form, and it provides as well terms of any order in the effective current expansion.

 Finally, we notice that the $Y_{ij}$ expression appears in a geometrical context as the component of the $R_{i+j+}$ Riemann tensor.
In terms of $R_{i+j+}$ we have
\beq\label{SecC14}
\mE^{(R)}_{ij}\,=\,-\,\D_{+}^{-2}R_{i+j+}\,,\qquad G^{(R)}_{ij}\,=\,-\,R_{i+j+}\,,\qquad
W^{(R)}_{ij}\,=\,\Big[P\exp\Big(-\,\D_{+}^{-2}R_{k+l+}\Big)\Big]_{ij}\,.
\eeq
The three forms of the generator are related by
\beq\label{SecC15}
\mE^{(K)}_{ij}\,=\,\mE^{(Y)}_{ij}\Big|_{h_{+i}=0}\,,
\qquad
\mE^{(R)}_{ij}\,=\,\mE^{(Y)}_{ij}\,+\,\tfrac12\,h_{ij}\,.
\eeq
We discuss the purely geometrical construction of the ordered exponential further in Section~\ref{SecF},
here we notice that the $h_{ij}$ is the least contributing in the high energy limit we consider and in general it should not be accounted for in the high energy
Regge limit we discuss.

\section{Effective current verification}\label{SecD}

 The simplest variant for the current we can use, which nevertheless provides the leading order contribution, is the $W_{ij}\,=\,W^{(K)}_{ij}$ operator,
see \eq{SecC9}-\eq{SecC10} definitions.
So, we take $h_{+i}=0$ and work with the three objects
\beq\label{SecD1}
K_{ij}\,=\,\D_i\D_j h_{++}\,,\qquad
G_{ij}\,=\,\frac{1}{2}\,K_{ij}\,,\qquad
\hat{G}_{ij}\,=\,\D_{+}^{-1}G_{ij}\,=\,\frac{1}{2}\,\D_{+}^{-1}K_{ij}\,,
\eeq
see \eq{SecC9}, \eq{SecC10} and \eq{SecC2}, so that
\beq\label{SecD2}
\D_{+}\hat{G}_{ij}\,=\,G_{ij}\,,\qquad
G_{ii}\,=\,\frac{1}{2}\,\D_{\bperp}^{2}h_{++}\,,\qquad
\hat{G}_{ij}\,=\,-\,\D_i X_{+j}\,,
\eeq
the last relation following from \eq{SecB3} at $h_{+i}\,=\,0$. The known second order answer of
\eq{SecB6} which we have to compare with is
\beq\label{SecD3}
j_{++}\,-\,h_{++}\,=\,X_{+i}X_{+i}\,,\qquad
X_{+i}\,=\,-\,\frac{1}{2}\,\D_i\D_{+}^{-1}h_{++}\,.
\eeq
The iterative solution \eq{SecC7} of \eq{SecC4} in turn provides
\beq\label{SecD4}
W_{ij}\,=\,\delta_{ij}\,+\,\Le \D_{+}^{-1}\hat{G}\Ra_{ij}\,+\,
\Big[\D_{+}^{-1}\Le \hat{G}\,\D_{+}^{-1}\hat{G}\Ra\Big]_{ij}\,+\,
\Big[\D_{+}^{-1}\Le \hat{G}\,\D_{+}^{-1}\Le \hat{G}\,\D_{+}^{-1}\hat{G}\Ra\Ra\Big]_{ij}\,+\,\dots\,,
\eeq
with the $P$ ordering of \eq{SecC3} built into the nesting of the $\D_{+}^{-1}$ operators.
Using \eq{SecC8} we obtain to the first order precision
\beq\label{SecD5}
\mO_{++}\big|_{h}\,=\,c\,G_{ii}\,=\,\frac{1}{2}\,c\,\D_{\bperp}^{2}h_{++}\,=\,\D_{\bperp}^{2}h_{++}
\,\rightarrow\,c\,=\,2\,,
\eeq
i.e. it fixes the \eq{SecC6} normalization. To the second order the two terms of \eq{SecC8}
combine, by $\D_{+}\hat{G}\,=\,G$ and the product rule of \eq{AppA2} of Appendix~\ref{AppA} into a single total derivative
\beq\label{SecD6}
\mO_{++}\big|_{h^{2}}\,=\,
c\,\Le G_{ik}\Le \D_{+}^{-1}\hat{G}\Ra_{ki}\,+\,\hat{G}_{ik}\hat{G}_{ki}\Ra\,=\,
c\,\D_{+}\Big[\hat{G}_{ik}\Le \D_{+}^{-1}\hat{G}\Ra_{ki}\Big]\,=\,
2\,\D_{+}\Big[\hat{G}_{ik}\Le \D_{+}^{-1}\hat{G}\Ra_{ki}\Big]\,.
\eeq
In terms of $X_{+i}$, by the last relation of \eq{SecD2}
\beq\label{SecD7}
\mO_{++}\big|_{h^{2}}\,=\,2\,\D_{+}\Big[\big(\D_i X_{+k}\big)\,\D_{+}^{-1}\big(\D_k X_{+i}\big)\Big]\,,
\eeq
or, carrying out the derivative and using $\D_{+}X_{+k}\,=\,-\tfrac12\,\D_k h_{++}$ together
with the symmetry of $\D_i X_{+k}$ in $i$ and $k$,
\beq\label{SecD8}
\mO_{++}\big|_{h^{2}}\,=\,2\,\big(\D_i \D_{+}X_{+k}\big)\,\D_{+}^{-1}\big(\D_k X_{+i}\big)\,+\,
2\,\big(\D_i X_{+k}\big)\big(\D_i X_{+k}\big)\,.
\eeq
Comparing now with \eq{SecB6}, where the first order terms cancel by \eq{SecD5}, and using \eq{AppA4},
\beq\label{SecD801}
\D_{\bperp}^{2}\big(X_{+k}X_{+k}\big)\,=\,2\big(\D_i X_{+k}\big)^{2}\,+\,
2\,X_{+k}\D_{\bperp}^{2}X_{+k}\,,
\eeq
we obtain from \eq{SecD8} and \eq{SecD801} for the mismatch
$\delta\mO_{++}\,\equiv\,\mO_{++}\big|_{h^{2}}\,-\,\D_{\bperp}^{2}\big(j_{++}-h_{++}\big)$
\beq\label{SecD9}
\delta\mO_{++}\,=\,
2\,\big(\D_i \D_{+}X_{+k}\big)\,\D_{+}^{-1}\big(\D_k X_{+i}\big)\,-\,
2\,X_{+k}\,\D_{\bperp}^{2}X_{+k}\,.
\eeq
The third order term of \eq{SecC8} gives correspondingly
\beq\label{SecD10}
\mO_{++}\big|_{h^{3}}\,=\,2\,G_{ik}\Big[\D_{+}^{-1}\Le \hat{G}\,\D_{+}^{-1}\hat{G}\Ra\Big]_{ki}
\,+\,2\,\hat{G}_{ik}\,\hat{G}_{kl}\,\Le \D_{+}^{-1}\hat{G}\Ra_{li}\,,
\eeq
which in terms of $X_{+i}$ reads
\beq\label{SecD11}
\mO_{++}\big|_{h^{3}}\,=\,-\,2\,\big(\D_i\D_{+}X_{+k}\big)\Big[\D_{+}^{-1}\Le \D_k X_{+l}\;
\D_{+}^{-1}\D_l X_{+i}\Ra\Big]\,-\,
2\,\big(\D_i X_{+k}\big)\big(\D_k X_{+l}\big)\,\D_{+}^{-1}\big(\D_l X_{+i}\big)\,,
\eeq
and was not obtained in \cite{MyGrav}. In general, the closed form of the operator
\beq\label{SecD12}
\mO_{++}\,=\,2\,\D_{+}\Big[\hat{G}_{ik}W_{ki}\Big]\,=\,
2\,\Le G_{ik}\,+\,\hat{G}_{il}\hat{G}_{lk}\Ra W_{ki}\,,
\qquad \D_{+}W_{ij}\,=\,\hat{G}_{ik}\,W_{kj}\,,
\eeq
allows one to calculate the operator to any requested perturbative order.

\section{The forms of the effective action}\label{SecE}

 The main purpose of the proposed formalism is the construction of a high energy gravitational effective action of Lipatov type. In this Section, therefore, we write 
the effective action which describes the interaction of Reggeized gravitons with regular ones using the operators we introduced. The form of the action is the same as 
found in \cite{MyGrav} and we use here the simplest \eq{SecC9} variant for the operators. So, the part which describes the interaction between the Reggeized gravitons and regular ones is 
\beq\label{SecE1}
L_{int}=\frac{1}{2\kappa}\,\mB_{--}\;\mO_{++}
=\frac{1}{2\kappa}\,\mB_{--}\Big[\D_{\bperp}^{2}h_{++}\,+\,2\,\D_{+}\Le \hat{G}^{(+)}_{ik}\Le W_{ki}^{(+)}-\delta_{ki}\Ra\Ra\Big]\,.
\eeq
The whole effective gravitational action is then
\beqar
L_{\rm eff}\,&=&\,L_{EH}(\eta,h)\;+\;
\frac{1}{2\kappa}\,\mB_{--}\Big[\D_{\bperp}^{2}h_{++}\,+\,2\,\D_{+}\Le \hat{G}^{(+)}_{ik}\big(W^{(+)}_{ki}-\delta_{ki}\big)\Ra\Big]\,+\,
\nonumber\\
&+&\,
\frac{1}{2\kappa}\,\mB_{++}\Big[\D_{\bperp}^{2}h_{--}\,+\,2\,\D_{-}\Le \hat{G}^{(-)}_{ik}\big(W^{(-)}_{ki}-\delta_{ki}\big)\Ra\Big]
\;-\;\frac{1}{\kappa}\,\mB_{++}\D_{\bperp}^{2}\mB_{--}\,,
\label{SecE2}
\eeqar
see for definitions \cite{MyGrav}; the superscript $\pm$ separates the quantities defined in terms of the $h_{++}$ and $h_{--}$ graviton components correspondingly.
In the expression we factorized out the $h_{\pm\pm}$ parts of the effective currents, so solving LO equations for the $h_{\pm\pm}$ fields and inserting back the
solutions, we obtain a RFT (Regge Field Theory) variant of the action written in terms of the Reggeized gravitons:
\beq\label{SecE3}
L_{\rm eff}\,=\,L_{EH}(\eta,\mB)\,+\,\frac{1}{\kappa}\sum_{\pm}\mB_{\mp\mp}\;
\D_{\pm}\Big(\hat{G}^{(\pm)}_{ik}\big[\mB\big]
\Big(W^{(\pm)}_{ki}\big[\mB\big]\,-\,\delta_{ki}\Big)\Big)\,.
\eeq
For example, expanding the expression, and writing only the plus line, we obtain triple and fourth orders of the interaction vertices of the Reggeized gravitons:
\beqar\label{SecE4}
L^{(3)}\,&=&\,\frac{1}{\kappa}\,\mB_{--}\,
\D_{+}\Big[\hat{G}_{ik}\Le \D_{+}^{-1}\hat{G}\Ra_{ki}\Big]\,;
\\
L^{(4)}\,&=&\,\frac{1}{\kappa}\,\mB_{--}\,
\D_{+}\Big[\hat{G}_{ik}\Big(\D_{+}^{-1}\Le \hat{G}\,\D_{+}^{-1}\hat{G}\Ra\Big)_{ki}\Big]\,;
\label{SecE5}
\eeqar
see \eq{SecD10} expression, here $\hat{G}_{ij}\,=\,\tfrac12\,\D_{+}^{-1}\D_i\D_j\mB_{++}$ as defined above with the $h_{++}$ field replaced by the Reggeon one. The \eq{SecE5} vertex is the new one, it was not derived in \cite{MyGrav}.

 As was mentioned above, see derivations in \cite{OurZubkov1,Our11,OurZubkov2}, the Reggeon fields appear in the action as the auxiliary ones and can be integrated out.
In this formulation what we describe is an effective action of two interacting P-ordered operators whose interaction is defined through the Lagrangian of
regular fields at some given local rapidity value. So, let us write the classical equations of motion for the $\mB$ fields. We have
\beq\label{SecE6}
L_{\mB}\,=\,-\frac{1}{\kappa}\,\mB_{++}\D_{\bperp}^{2}\mB_{--}
+\frac{1}{2\kappa}\Le \mB_{--}\mO_{++}+\mB_{++}\mO_{--}\Ra\,,
\eeq
and the equations of motion read
\beq\label{SecE7}
\frac{\delta L}{\delta \mB_{++}}\,=\,0\;\rightarrow\; \mB_{--}\,=\,\tfrac12\D_{\bperp}^{-2}\mO_{--}\,,
\qquad
\frac{\delta L}{\delta \mB_{--}}\,=\,0\;\rightarrow\; \mB_{++}\,=\,\tfrac12\D_{\bperp}^{-2}\mO_{++}\,.
\eeq
Substituting the solutions back we obtain
\beqar
L_{\rm eff}\,&=&L_{EH}(\eta,h)\,-\,\frac{1}{4\kappa}\,\mO_{++}\D_{\bperp}^{-2}\mO_{--}\,+\,\frac{1}{2\kappa}\,\mO_{++}\D_{\bperp}^{-2}\mO_{--}\,=\,
\nonumber
\\
&=&
L_{EH}(\eta,h)\,+\,\frac{1}{4\kappa}\;\mO_{++}\,\D_{\bperp}^{-2}\,\mO_{--}\,=\,
L_{EH}(\eta,h)\,+\,\frac{1}{\kappa}\Big(\D_{+}\big[\hat{G}^{(+)}_{ik}W^{(+)}_{ki}\big]\Big)\;
\D_{\bperp}^{-2}\;\Big(\D_{-}\big[\hat{G}^{(-)}_{jl}W^{(-)}_{lj}\big]\Big)\,,
\label{SecE8}
\eeqar
which is a variant of the gravitational high energy effective action written directly in terms of the regular gravitons.

 We also notice that the interaction term of \eq{SecE8} is purely transverse. By the kinematic constraints the two
factors depend on one light cone coordinate each, $\mO_{++}\,=\,\mO_{++}(x^{+},x_{\bperp})$ and
$\mO_{--}\,=\,\mO_{--}(x^{-},x_{\bperp})$, each of them is a total derivative in that
coordinate, and $\D_{\bperp}^{-2}$ commutes with both longitudinal integrations. Therefore
\beq\label{SecE9}
\frac{1}{4\kappa}\int\! d^{4}x\;\,\mO_{++}\,\D_{\bperp}^{-2}\,\mO_{--}\,=\,
\frac{1}{\kappa}\int\! d^{2}x_{\bperp}\,
\Big[\hat{G}^{(+)}_{ik}W^{(+)}_{ki}\Big]_{x^{+}=-\infty}^{x^{+}=+\infty}\;
\D_{\bperp}^{-2}\;
\Big[\hat{G}^{(-)}_{jl}W^{(-)}_{lj}\Big]_{x^{-}=-\infty}^{x^{-}=+\infty}\,,
\eeq
a two dimensional integral of two boundary values. With the symmetric definition of
$\D_{\pm}^{-1}$ assumed after \eq{SecC8} one has
$\hat{G}\big|_{x^{+}\to\pm\infty}\,=\,\pm\tfrac12\,\bar{G}$ with
$\bar{G}_{ij}(x_{\bperp})\,=\,\int_{-\infty}^{+\infty}\!dx_{1}^{+}\,G_{ij}(x_{1}^{+},x_{\bperp})$
independent of $x^{+}$, and $W\big|_{x^{+}\to-\infty}\,=\,\bone$, so that each
boundary value takes the closed form
\beq\label{SecE10}
\int\! dx^{+}\,\mO_{++}\,=\,2\,\Big[\hat{G}_{ik}W_{ki}\Big]_{-\infty}^{+\infty}\,=\,
\tr\Big[\bar{G}\,\Le W(+\infty)\,+\,\bone\Ra\Big]\,,
\eeq
with $W(+\infty)\,=\,P\exp\int_{-\infty}^{+\infty}\!dx^{+}\hat{\bG}$ the complete ordered
exponential of the line. At the leading order $W(+\infty)\,=\,\bone$ and \eq{SecE10} returns
$2\,\tr\bar{G}\,=\,\int dx^{+}\D_{\bperp}^{2}h_{++}$, the Born vertex. The whole interaction
depends on the fields therefore only through $\bar{G}$ and the complete ordered exponential of
each line, contracted across the transverse plane.
We underline that \eq{SecE10} requires a regularization. Taken literally over the whole light
cone the connection $\hat{G}_{ij}$ does not vanish at the ends and the two prescriptions of
\eq{SecC8} fail in complementary ways. With the retarded primitive $\hat{G}\rightarrow\bar{G}$ at
the upper end, its integral grows linearly and the ordered exponential $W(+\infty)$ diverges
already for the shock wave profile, while $W(x^{+})$ at finite $x^{+}$ stays finite. With the
symmetric one $\hat{G}\rightarrow\pm\tfrac12\bar{G}$ at the two ends and
$\int_{-L}^{x^{+}}dx_{1}^{+}\hat{G}=\tfrac12(x^{+}-L)\bar{G}-\int dx_{1}^{+}x_{1}^{+}G$ beyond the
support of $G$, so that $W(+\infty)=\exp\big(-\int dx^{+}x^{+}G\big)$ is finite for commuting
generators and depends on the origin, while $W(x^{+})$ at finite $x^{+}$ diverges with the cutoff,
the normalization $W|_{-\infty}=\bone$ of \eq{SecC4} being imposed where $\hat{G}$ does not vanish.
The expression is therefore understood with some regularization introduced.

\section{Geometry of the ordered exponential}\label{SecF}

 We return now to \eq{SecC14} expressions. Discussing gravity we talk about geometry so the observation made is simple, the auxiliary variables \eq{SecB3}-\eq{SecB5} sit in the Riemann tensor linearized over flat background
\beq\label{SecF1}
R_{\mu\nu\rho\sigma}\,=\,\tfrac12\Le \D_\nu\D_\rho h_{\mu\sigma}+\D_\mu\D_\sigma h_{\nu\rho}\,-\,\D_\nu\D_\sigma h_{\mu\rho}\,-\,\D_\mu\D_\rho h_{\nu\sigma}\Ra\,.
\eeq
In light cone variables with $\mu\,=\,i$, $\nu\,=\,+$, $\rho\,=\,j$, $\sigma=+$, the four terms become
\beq\label{SecF2}
\D_\nu\D_\rho h_{\mu\sigma}\to\D_{+}\D_j h_{i+}\,,\quad
\D_\mu\D_\sigma h_{\nu\rho}\to\D_i\D_{+}h_{+j}\,,\quad
\D_\nu\D_\sigma h_{\mu\rho}\to\D_{+}\D_{+}h_{ij}\,,\quad
\D_\mu\D_\rho h_{\nu\sigma}\to\D_i\D_j h_{++},
\eeq
so that
\beq\label{SecF3}
R_{i+j+}\,=\,\tfrac12\Le \D_{+}\D_j h_{i+}\,+\,\D_{+}\D_i h_{+j}\,-\,\D_{+}^{2}h_{ij}\,-\,\D_i\D_j h_{++}\Ra\,.
\eeq
Now multiply by $2$ and act with $\D_{+}^{-2}$, term by term
\beqar
2\D_{+}^{-2}R_{i+j+}\,&=&\,\D_{+}^{-2}\D_{+}\Le \D_j h_{i+}\,+\,\D_i h_{+j}\Ra\,-\,\D_{+}^{-2}\D_{+}^{2}h_{ij}\,-\,\D_{+}^{-2}\D_i\D_j h_{++}\,=\,
\nonumber
\\
&=&\,
\D_{+}^{-1}\Le \D_i h_{+j}\,+\,\D_j h_{+i}\Ra\,-\,\D_{+}^{-2}\D_i\D_j h_{++}\,-\,h_{ij}\, .
\label{SecF4}
\eeqar
We see that
\beq\label{SecF5}
Y_{ij}\,=\,2\,\D_{+}^{-2}R_{i+j+}\,+\,h_{ij}\,,\qquad\,
\mE^{(Y)}_{ij}\,=\,-\,\tfrac12 Y_{ij}\,=\,-\,\D_{+}^{-2}R_{i+j+}\,-\,\tfrac12 h_{ij}\,=\,\mE^{(R)}_{ij}-\tfrac12 h_{ij}\,,
\eeq
as written in \eq{SecC15}. In the gauge with $h_{+i}\,=\,h_{ij}\,=\,0$ only the last term of \eq{SecF3} survives,
\beq\label{SecF6}
R_{i+j+}\,=\,-\tfrac12\,\D_i\D_j h_{++}\,,\qquad
\mE^{(K)}_{ij}\,=\,\tfrac12\D_{+}^{-2}\D_i\D_j h_{++}\,=\,-\,\D_{+}^{-2}R_{i+j+}=\mE^{(R)}_{ij}\,.
\eeq
With $G_{ij}\,=\,-\,R_{i+j+}$ the connection of the ordered exponential is minus the first light cone
primitive of the tidal matrix, $\hat{G}^{(R)}_{ij}\,=\,-\,\D_{+}^{-1}R_{i+j+}$, and the
expressions \eq{SecC4} and \eq{SecC6} read
\beq\label{SecF7}
\D_{+}W^{(R)}_{ij}\,=\,-\,\Le \D_{+}^{-1}R\Ra_{i+k+}\,W^{(R)}_{kj}\,,\qquad
\mO_{++}\,=\,-\,2\,\D_{+}\Big[\Le \D_{+}^{-1}R\Ra_{i+k+}\,W^{(R)}_{ki}\Big]\,.
\eeq
The second order form \eq{SecC5} of the same equation reads
\beq\label{SecF8}
\D_{+}^{2}W^{(R)}_{ij}\,=\,\Le \D_{+}\hat{G}^{(R)}_{ik}\,+\,
\hat{G}^{(R)}_{il}\hat{G}^{(R)}_{lk}\Ra\,W^{(R)}_{kj}\,,\qquad
\D_{+}\hat{G}^{(R)}_{ik}\,+\,\hat{G}^{(R)}_{il}\hat{G}^{(R)}_{lk}\,=\,
-\,R_{i+k+}\,+\,O(h^{2})\,,
\eeq
so that the effective tidal matrix of $W^{(R)}_{ij}$ is minus the true one at the linear order only,
where it reduces to \eq{SecF3}; see also \eq{SecC14}.

Two objects have to be kept apart from here on. The first is $W^{(R)}_{ij}$ of \eq{SecF7}, the
ordered exponential of $\hat{G}^{(R)}_{ij}=-\D_{+}^{-1}R_{i+j+}$. It is the covariant writing of
$W_{ij}$ of \eq{SecC3} and it is the object that carries the effective current. The second is
the Jacobi propagator $W^{\rm J}_{ij}$ of the null congruence, generated by the deformation
matrix $\hat{s}_{ij}$ which solves the Riccati equation with the curvature itself,
\beq\label{SecF801}
\D_{+}W^{\rm J}_{ij}\,=\,\hat{s}_{ik}\,W^{\rm J}_{kj}\,,\qquad
\D_{+}\hat{s}_{ik}\,+\,\hat{s}_{il}\hat{s}_{lk}\,=\,\mR_{ik}\,=\,R_{i+k+}\,,\qquad
\D_{+}^{2}W^{\rm J}_{ij}\,=\,\mR_{ik}\,W^{\rm J}_{kj}\,.
\eeq
This is \eq{SecG9} in the light cone frame. The relative sign of the two generators depends on the
position of a single transverse index in the expressions. With $\eta_{ij}=-\delta_{ij}$ of \eq{SecB9} the raising adds
a minus, and the convention adopted after \eq{SecB10}, which keeps all transverse indices down,
hides this. Taking the arrays as they are written,
\beq\label{SecF802}
\hat{G}^{(R)}_{ij}\,=\,-\,\D_{+}^{-1}R_{i+j+}\,,\qquad
\eta^{jk}\,\hat{G}^{(R)}_{ik}\,=\,\D_{+}^{-1}R_{i+j+}\,=\,\hat{s}_{ij}\,+\,O(h^{2})\,,\qquad
\bW^{(R)}\,=\,\Le \bW^{\rm J}\Ra^{-1}\,+\,O(h^{2})\,,
\eeq
we see that with the index raised the Jacobi sign is positive again. The two index positions are two
different transports. The vector index carries the deviation vector $\xi^{i}$ and the covector one
carries the deviation form $\xi_{i}$, the preservation of $\xi_{i}\xi^{i}$ makes the two inverse
transposed to each other, which for a symmetric generator is the opposite sign; the transpose itself
is invisible at the first order, where $\hat{s}$ is symmetric. The choice made in \eq{SecC10} is
therefore the choice of the form transport for the effective current. The two separate beyond the
first order, as \eq{SecI11} shows in closed form for the shock wave. The inversion in
\eq{SecF802} is written with the retarded primitive, the Riccati equation \eq{SecF801} being
integrated from $\hat{s}(-\infty)=0$. With the symmetric $\D_{+}^{-1}$ of \eq{SecC8} the two
primitives differ by a constant, $\hat{S}_{S}=\hat{S}_{R}+\tfrac12\bar{\mR}$, and
$\bW\Le \bW^{\rm J}\Ra^{T}=\bone+\tfrac12\bar{\mR}\,(x^{+}-x^{+}_{0})+O(h^{2})$, a defect of the
first order growing with the length of the line. The defect is traceless where
$\tr\bar{\mR}\propto\int dx^{+}T_{++}$ vanishes, so away from the matter
$\det\bW\,\det\bW^{\rm J}=1+O(h^{2})$ in either prescription and the statements built on the
determinant alone are not affected. The current of \eq{SecF7} is built on the first; the geometry of the
congruence, the optical scalars, the caustics and the van Vleck determinant belong to the
second. The symbol $\mR$ denotes the curvature projection throughout and nothing else.

To LO precision with $W^{(R)}_{ki}\,=\,\delta_{ki}$ we obtain in turn
\beq\label{SecF9}
\mO_{++}\big|_{LO}\,=\,-2\,\delta^{ik}R_{i+k+}\,=\,2\,\eta^{ik}R_{i+k+}\,=\,2\,R_{++}\,.
\eeq
Rewriting now the effective action expressions we obtain the following answers for the different terms of the effective action. 

 The part of the Lagrangian which describes the interactions of Reggeons and gravitons in terms of the Riemann tensor has the following form now
\beqar
L_{int}\,&=&\,\frac{1}{2\kappa}\,\mB_{--}\,\mO_{++}\,=\,
-\,\frac{1}{\kappa}\;\mB_{--}\;\D_{+}\Big[\Le \D_{+}^{-1}R\Ra_{i+k+}W^{(R)}_{ki}\Big]\,=\,
\nonumber
\\
&=&\,\frac{1}{\kappa}\;\mB_{--}\,R_{++}\,-\,
\frac{1}{\kappa}\;\mB_{--}\;\D_{+}\Big[\Le \D_{+}^{-1}R\Ra_{i+k+}\Le W^{(R)}_{ki}\,-\,\delta_{ki}\Ra\Big]\,.
\label{SecF10}
\eeqar
It is interesting to notice that introducing
\beq\label{SecF11}
\mR^{\mB}_{i-j-}\,\equiv\,-\tfrac12\,\D_i\D_j\mB_{--},\qquad
\D_i^{2}\mB_{--}\,=\,-2\,\delta^{ij}\mR^{\mB}_{i-j-}\,=\,2\mR^{\mB}_{--}\,=\,2\,\eta^{ij}\mR^{\mB}_{i-j-}\,
\eeq
aka the curvature of the Reggeized graviton, the same interaction term we can write as
\beq\label{SecF12}
L_{int}\,=\,\frac{1}{2\kappa}\,\D_i^{2}\mB_{--}\,\D_{\bperp}^{-2}\mO_{++}\,=\,
\frac{1}{\kappa}\,\mR^{\mB}_{--}\,\D_{\bperp}^{-2}\mO_{++}\,=\,
-\,\frac{2}{\kappa}\,\mR^{\mB}_{--}\,\D_{\bperp}^{-2}\,
\D_{+}\Big[\Le \D_{+}^{-1}R\Ra_{i+k+}\,W^{(R)}_{ki}\Big]\,,
\eeq
or as
\beq\label{SecF13}
L_{int}\big|_{LO}\,=\,\frac{2}{\kappa}\,\mR^{\mB}_{--}\,\D_{\bperp}^{-2}\,R_{++}\,
\eeq
to leading order precision. The same interaction term written in terms of the operators, the \eq{SecE8} expression, now reads
\beq\label{SecF14}
L_{\rm int}\,=\,\frac{1}{4\kappa}\,\mO_{++}\D_{\bperp}^{-2}\mO_{--}
\,=\,\frac{1}{\kappa}\Big(\D_{+}\big[\Le \D_{+}^{-1}R\Ra_{i+k+}\,W^{(R,+)}_{ki}\big]\Big)\;\D_{\bperp}^{-2}\;
\Big(\D_{-}\big[\Le \D_{-}^{-1}R\Ra_{j-l-}\,W^{(R,-)}_{lj}\big]\Big)\,,
\eeq
where $W^{(R,+)}$ is ordered in $x^{+}$ with generator $R_{i+j+}$, and $W^{(R,-)}$ is ordered
in $x^{-}$ with generator $R_{i-j-}$. At leading order, by \eq{SecF9},
\beq\label{SecF15}
L_{\rm int}\,\to\,\frac{1}{\kappa}\,R_{++}\,\D_{\bperp}^{-2}\,R_{--}\,=\,
\tfrac{\kappa}{4}\,T_{++}\,\D_{\bperp}^{-2}\,T_{--}\,,
\eeq
in agreement with the Newtonian limit. Finally we write as well the RFT form of the action, it is
\beq\label{SecF16}
L_{\rm eff}\,=\,L_{EH}(\eta,\mB)\,-\,\frac{1}{\kappa}\sum_{\pm}\mB_{\mp\mp}\;
\D_{\pm}\Big[\Le \D_{\pm}^{-1}R^{(\pm)}\Ra_{i\pm k\pm}\big[\mB\big]
\Le W^{(R,\pm)}_{ki}\big[\mB\big]-\delta_{ki}\Ra\Big]\,,
\eeq
with the curvature built from the Reggeon field itself.

 We notice now that the interaction term \eq{SecF14} is purely transverse as well:
\beq\label{SecF17}
\int\! d^{4}x\,L_{\rm int}\,=\,\frac{1}{\kappa}\int\! d^{2}x_{\bperp}\,
\Big[\Le \D_{+}^{-1}R\Ra_{i+k+}W^{(R,+)}_{ki}\Big]_{x^{+}=-\infty}^{x^{+}=+\infty}\;
\D_{\bperp}^{-2}\;
\Big[\Le \D_{-}^{-1}R\Ra_{j-l-}W^{(R,-)}_{lj}\Big]_{x^{-}=-\infty}^{x^{-}=+\infty}\,.
\eeq
Performing the calculations similarly to what was done for \eq{SecE10}, we obtain
\beq\label{SecF18}
\int\! d^{4}x\,L_{\rm int}\,=\,\frac{1}{4\kappa}\int\! d^{2}x_{\bperp}\,
\tr\Big[\bar{R}^{(+)}\Le W^{(R,+)}(+\infty)\,+\,\bone\Ra\Big]\;\D_{\bperp}^{-2}\;
\tr\Big[\bar{R}^{(-)}\Le W^{(R,-)}(+\infty)\,+\,\bone\Ra\Big]\,,
\eeq
where
\beq\label{SecF1801}
\bar{R}^{(\pm)}_{ij}(x_{\bperp})\,=\,\int_{-\infty}^{+\infty}\!dx_{1}^{\pm}\,
R_{i\pm j\pm}(x_{1}^{\pm},x_{\bperp})
\eeq
are independent of $x^{\pm}$. 
We underline again that \eq{SecF18} requires a regularization for the same reason as
\eq{SecE10} does. Similarly, the interaction part of \eq{SecF16} RFT formulation is also purely transverse. Introducing transverse Reggeon field 
similarly to done in \eq{SecE6}-\eq{SecE7} through the corresponding longitudinal we obtain
\beq\label{SecF19}
\int\! d^{4}x\,L_{int}\,=\,-\,\frac{1}{\kappa}\int\! d^{2}x_{\bperp}\sum_{\pm}
\bar{\mB}_{\mp\mp}\,
\Big[\Le \D_{\pm}^{-1}R\Ra_{i\pm k\pm}\Le W^{(R,\pm)}_{ki}\,-\,\delta_{ki}\Ra\Big]_{x^{\pm}=-\infty}^{x^{\pm}=+\infty}\,,
\qquad
\bar{\mB}_{\mp\mp}(x_{\bperp})\,=\,\int_{-\infty}^{+\infty}\!dx_{1}^{\mp}\,
\mB_{\mp\mp}(x_{1}^{\mp},x_{\bperp})\,.
\eeq
With the symmetric definition of $\D_{\pm}^{-1}$ one has
\beq\label{SecF1901}
\Le \D_{\pm}^{-1}R\Ra\big|_{\pm\infty}\,=\,\pm\tfrac12\bar{R}\,,\qquad W\big|_{-\infty}\,=\,\bone\,,
\eeq
with
\beq\label{SecF1902}
\qquad\,
\big[\Le \D_{\pm}^{-1}R\Ra_{i\pm k\pm}\Le W_{ki}-\delta_{ki}\Ra\big]_{-\infty}^{+\infty}\,=\,
\tfrac12\tr\big[\bar{R}\Le W(+\infty)-\bone\Ra\big]\,,
\eeq
then the interaction takes the closed form
\beq\label{SecF20}
\int\! d^{4}x\,L_{int}\,=\,-\,\frac{1}{2\kappa}\int\! d^{2}x_{\bperp}\sum_{\pm}
\bar{\mB}_{\mp\mp}\;\tr\Big[\bar{R}^{(\pm)}\Le W^{(R,\pm)}(+\infty)\,-\,\bone\Ra\Big]\,,
\qquad
\bar{R}^{(\pm)}_{ij}(x_{\bperp})\,=\,\int_{-\infty}^{+\infty}\!dx_{1}^{\pm}\,
R^{(\pm)}_{i\pm j\pm}\big[\mB\big](x_{1}^{\pm},x_{\bperp})\,.
\eeq
Unlike \eq{SecF14} the expression is local in the transverse plane, since the Reggeon fields
are not integrated out here and no transverse propagator has been generated yet. The
subtraction of $\delta_{ki}$ removes the free term, so that the first contribution to
\eq{SecF20} is the three Reggeon vertex. Only the interaction part reduces in this way, the
$L_{EH}(\eta,\mB)$ term of \eq{SecF16} is not a total longitudinal derivative.

 At the end of the Section we note that the interaction term of the action is now formulated in the way that 
the order by order procedure of \eq{SecB2} is not requested anymore. Whereas the $Y_{ij}$ is invariant under $\varepsilon_{+}$
only, the $R_{i+j+}$ is invariant under all linearized diffeomorphisms. The connection
$\hat{G}^{(R)}_{ij}\,=\,-\,\D_{+}^{-1}R_{i+j+}$ inherits the invariance and so does the tidal
matrix $\mR_{ik}$ of \eq{SecF801}, being built from invariants. The statement holds for the gauge
parameters which fall off at the ends of the line, the boundary condition carried by
$\D_{+}^{-1}$.

\section{Covariant form and the optical parameters of the ordered exponential}\label{SecG}

 As defined, the operator $\D_{+}^{-1}$ is not a tensor, so \eq{SecF7} is still connected to the
light cone gauge and to a fixed null direction. With the geometrical definition of the interactions, now we can consider a covariant
description of the following process. We take a single null line in a given external field: the metric is
an input, there is no back reaction, no second source and no Reggeon. The dynamical problem,
the coupling of $\mO_{\pm\pm}$ to the Reggeon field and the convolution of the two lines across
the screen, is that of Sections \ref{SecE}, \ref{SecF} and \ref{SecH}. The descriptions intersect in
\eq{SecG23}, where the current which enters the action is expressed through the invariants of
the ray bundle. To some extent, therefore, this description is similar to the \cite{BalOper} description of high energy processes.

 Now, let $k^{\mu}$ be the null tangent of the null geodesic followed by the fast particle of the plus
line
\beq\label{SecG1}
k^{\mu}k_{\mu}\,=\,0\,,\qquad k^{\nu}\nabla_\nu k^{\mu}\,=\,0
\eeq
with affine parameter $\lambda$, and
let $\bar k^{\mu}$ be the second null vector of the Regge frame, $k\cdot\bar k\neq0$, tangent
to the minus line. Let $e_{A}^{\mu}$, $A=1,2$, be a Sachs dyad on the screen orthogonal to
both, parallel transported along the congruence,
\beq\label{SecG2}
k^{\nu}\nabla_{\nu}e_{A}^{\mu}\,=\,0,\qquad e_A\cdot k\,=\,e_A\cdot\bar k\,=\,0,\qquad
e_A\cdot e_B\,=\,-\delta_{AB}\,.
\eeq
Define the optical tidal matrix of the plus line and its mirror,
\beq\label{SecG3}
\mR^{(+)}_{AB}\,=\,R_{\mu\nu\rho\sigma}\,e_{A}^{\mu}k^{\nu}e_{B}^{\rho}k^{\sigma}\,,
\qquad
\mR^{(-)}_{AB}\,=\,R_{\mu\nu\rho\sigma}\,e_{A}^{\mu}\bar k^{\nu}e_{B}^{\rho}\bar k^{\sigma}\,,
\eeq
which reduce to $R_{i+j+}$ and $R_{i-j-}$ in the light cone frame. Let $\hat{S}^{(\pm)}_{AB}$
be the covariant counterpart of the connection $\hat{G}_{ij}$ of \eq{SecC2}, defined by the
same first light cone primitive,
\beq\label{SecG301}
\frac{D\hat{S}^{(\pm)}_{AB}}{d\lambda_{\pm}}\,=\,-\,\mR^{(\pm)}_{AB}\,,\qquad
\hat{S}^{(\pm)}_{AB}\,=\,-\,\frac{1}{2}\int\! d\lambda'\;
\epsilon\big(\lambda_{\pm}-\lambda'\big)\,\mR^{(\pm)}_{AB}(\lambda')\,,\qquad
\hat{S}^{(\pm)}_{AB}\big|_{\lambda_{\pm}\to\pm\infty}\,=\,\mp\,\tfrac12\,\bar{\mR}^{(\pm)}_{AB}\,,\qquad
\eeq
with
\beq\label{SecG302}
\bar{\mR}_{AB}\,=\,\int_{-\infty}^{+\infty}\!\!d\lambda'\,\mR_{AB}(\lambda')\,.
\eeq
The primitive is the symmetric $\D_{\pm}^{-1}$ adopted after \eq{SecC8}, so that \eq{SecG301} is
\eq{SecC2} written covariantly and $\hat{S}$ does not vanish at either end of the line. We use further also the retarded
primitive which appears only in the shock wave example, see the remark after \eq{SecH702}.
Since $\mR_{AB}$ is symmetric by the pair symmetry of the Riemann tensor, $\hat{S}_{AB}$ is
symmetric identically. Then \eq{SecF7} is the covariant first order equation
\beq\label{SecG4}
\frac{DW^{(\pm)}_{AB}}{d\lambda_{\pm}}\,=\,\hat{S}^{(\pm)}_{AC}\,W^{(\pm)}_{CB},
\qquad W^{(\pm)}_{AB}\big|_{\lambda_\pm\to\,-\infty}\,=\,\delta_{AB}\,,
\qquad
\frac{D}{d\lambda_{+}}\,\equiv\,k^{\mu}\nabla_{\mu}\,,\quad
\frac{D}{d\lambda_{-}}\,\equiv\,\bar{k}^{\mu}\nabla_{\mu}\,,
\eeq
the derivative being taken along the corresponding null geodesic,
$k^{\mu}\,=\,dx^{\mu}/d\lambda_{+}$ and $\bar{k}^{\mu}\,=\,dx^{\mu}/d\lambda_{-}$. Acting on
the screen matrices it is the ordinary derivative,
\beq\label{SecG5}
\frac{DW^{(\pm)}_{AB}}{d\lambda_{\pm}}\,=\,\frac{dW^{(\pm)}_{AB}}{d\lambda_{\pm}}\,,
\qquad
\frac{D\hat{S}^{(\pm)}_{AB}}{d\lambda_{\pm}}\,=\,\frac{d\hat{S}^{(\pm)}_{AB}}{d\lambda_{\pm}}\,,
\eeq
since the dyad $e_{A}^{\mu}$ is parallel transported along the congruence by \eq{SecG2}, so
that the components of any screen matrix in this frame are scalars along the line.

By \eq{SecG301} the second order form of \eq{SecG4} is \eq{SecC5} written covariantly,
\beq\label{SecG401}
\frac{D^{2}W^{(\pm)}_{AB}}{d\lambda_{\pm}^{2}}\,=\,
\Le -\,\mR^{(\pm)}_{AC}\,+\,\hat{S}^{(\pm)}_{AD}\hat{S}^{(\pm)}_{DC}\Ra\,W^{(\pm)}_{CB}\,,
\eeq
the second structure being the one added in \eq{SecC5}.

 There is a second matrix we introduce here separately from $\hat{S}_{AB}$, the deformation
matrix of the ray bundle itself, defined geometrically as the screen projection of the gradient
of the null tangent
\beq\label{SecG6}
\hat{s}^{(+)}_{AB}\,=\,e_{A}^{\mu}\,e_{B}^{\nu}\,\nabla_{\nu}k_{\mu}\,,\qquad
\hat{s}^{(-)}_{AB}\,=\,e_{A}^{\mu}\,e_{B}^{\nu}\,\nabla_{\nu}\bar{k}_{\mu}\,.
\eeq
It is the object which governs the deviation of neighboring rays of the congruence: for a
deviation vector $\xi^{\mu}$ connecting two neighboring null geodesics one has
\beq\label{SecG601}
D\xi^{\mu}/d\lambda\,=\,\xi^{\nu}\nabla_{\nu}k^{\mu}
\eeq
which on the screen reads as
\beq\label{SecG602}
d\xi_{A}/d\lambda\,=\,\hat{s}_{AB}\xi_{B}\,,
\eeq
so that the map of $\xi_{B}(-\infty)$ into $\xi_{A}(\lambda)$ is generated by $\hat{s}$ and not
by $\hat{S}$; we denote it $W^{\rm J}_{AB}$, the object of \eq{SecF801}.

 We decompose now the generator \eq{SecG301} of our ordered exponential into its irreducible
parts,
\beq\label{SecG7}
\hat{S}_{AB}\,=\,\tfrac12\,\theta\,\delta_{AB}\,+\,\sigma_{AB}\,+\,\omega_{AB}\,,\qquad
\theta\,=\,\delta^{AB}\hat{S}_{AB}\,,\qquad
\sigma_{AB}\,=\,\hat{S}_{(AB)}\,-\,\tfrac12\,\theta\,\delta_{AB}\,,\qquad
\omega_{AB}\,=\,\hat{S}_{[AB]}\,,
\eeq
where $\theta$ is the trace of the connection,
\beq\label{SecG8}
\theta\,=\,\frac{d}{d\lambda}\,\ln\det W_{AB}\,;
\eeq
it is minus the expansion of the ray bundle, since $\det\bW$ is the ray density and not the area
element, see \eq{SecG27}. The expansion itself is $\theta_{s}$ of \eq{SecG903};
$\sigma_{AB}$ is the shear, symmetric and traceless, which deforms a circular bundle of rays
into an ellipse of the same area; and $\omega_{AB}$ is the twist, antisymmetric, which rotates
the bundle as a whole. Here the twist vanishes identically and not by a propagation argument:
$\hat{S}$ is the primitive \eq{SecG301} of $\mR_{AB}$, which is symmetric by the pair symmetry
of the Riemann tensor, so
\beq\label{SecG701}
\omega_{AB}\,\equiv\,0\,,\qquad
\frac{d\theta}{d\lambda}\,=\,-\,\mR_{AA}\,,\qquad
\frac{d\sigma_{AB}}{d\lambda}\,=\,-\,\mR^{TF}_{AB}\,,
\eeq
where
\beq\label{SecG702}
\theta\,=\,-\,\int^{\lambda}\!\!d\lambda'\,\mR_{AA}\,,\qquad
\sigma_{AB}\,=\,-\,\int^{\lambda}\!\!d\lambda'\,\mR^{TF}_{AB}\,,
\eeq
in closed form. The congruence is hypersurface orthogonal. At leading order this is visible
directly, since $\hat{S}_{AB}$ then reduces to $\hat{G}_{ij}\,=\,-\,\D_i X_{+j}$ of \eq{SecD2},
symmetric in $i$ and $j$ when $h_{+i}=0$.
We notice also that for \eq{SecG6} $\hat{s}$, \eq{SecF801} is the Riccati relation between this deformation matrix and the tidal
one,
\beq\label{SecG9}
\frac{D\hat{s}^{(\pm)}_{AB}}{d\lambda_{\pm}}\,+\,\hat{s}^{(\pm)}_{AC}\hat{s}^{(\pm)}_{CB}
\,=\,\mR^{(\pm)}_{AB}\,,
\qquad\rightarrow\qquad
\frac{D^{2}W^{{\rm J}(\pm)}_{AB}}{d\lambda_{\pm}^{2}}\,=\,\mR^{(\pm)}_{AC}\,W^{{\rm J}(\pm)}_{CB}\,,
\eeq
so that $W^{\rm J}_{AB}$ is the Jacobi propagator of the full tidal matrix. Comparing then \eq{SecG301} with
\eq{SecG9} we obtain that the two matrices are related by an ordinary perturbative series: 
\beq\label{SecG902}
\hat{s}_{AB}\,=\,-\,\hat{S}_{AB}\,-\,\int^{\lambda}\!\!d\lambda'\,
\Le \hat{S}\hat{S}\Ra_{AB}(\lambda')\,+\,\dots\,,
\eeq
with the retarded primitive as in \eq{SecF802}, each further term carrying one more power of
$\hat{S}$ and therefore one more power of $h$, since $\hat{S}=-\D_{+}^{-1}\mR$ is of the first
order in the field by \eq{SecF3}. The optical
scalars of the bundle follow from \eq{SecG7} by the same series. Using the two dimensional
identities \eq{SecG10}, $\hat{S}\hat{S}=\Le \tfrac14\theta^{2}+\tfrac12\sigma^{2}\Ra\bone+
\theta\sigma$, so that
\beq\label{SecG903}
\theta_{s}\,=\,-\,\theta\,-\,\int^{\lambda}\!\!d\lambda'\Le \tfrac12\,\theta^{2}\,+\,\sigma^{2}\Ra
\,+\,\dots\,,\qquad
\sigma_{s\,AB}\,=\,-\,\sigma_{AB}\,-\,\int^{\lambda}\!\!d\lambda'\;\theta\,\sigma_{AB}
\,+\,\dots\,,\qquad \omega_{s}\,=\,0\,,
\eeq
the first correction being of the second order in $h$, the next of the third, and so on. The
Raychaudhuri equation of the bundle is then not an independent input but a consequence of
\eq{SecG701} and \eq{SecG903},
\beq\label{SecG904}
\frac{d\theta_{s}}{d\lambda}\,+\,\tfrac12\,\theta_{s}^{2}\,+\,\sigma_{s}^{2}\,=\,
\Le \mR_{AA}\,-\,\tfrac12\theta^{2}\,-\,\sigma^{2}\Ra\,+\,\tfrac12\theta^{2}\,+\,\sigma^{2}\,=\,
\mR_{AA}\,+\,O(h^{3})\,,
\eeq
the quadratic terms being exactly what compensates the difference between the two matrices, the
$O(h^{3})$ coming from the truncation of \eq{SecG903}. The equation itself is exact, being the
trace of \eq{SecG9}, and only its perturbative verification is of that order.
Everything built on $\hat{s}$ describes the ray bundle; the current of the action is built on
$\hat{S}$ and $W$ through \eq{SecG4}. The two readings are inverse to each other at the first
order, $\hat{s}=-\hat{S}+O(h^{2})$ by \eq{SecF802}, which is what allows the bundle language to be
used for $W$ below with the dictionary inverted.
Only the value of $W$ is fixed at $\lambda\to-\infty$; its derivative there equals
$\hat{S}(-\infty)\,=\,\tfrac12\bar{\mR}_{AB}$ by \eq{SecG301}.
The second order form of \eq{SecG9} is a linear system for the pair
$(W^{\rm J},DW^{\rm J}/d\lambda)$ on a phase space of dimension four, written explicitly in
\eq{SecI18}. Its generator
\beq\label{SecG901}
\begin{pmatrix}0&\bone\\ \mR&0\end{pmatrix}
\eeq
lies in the algebra $\mathfrak{sp}(4,\mathbb{R})$ because $\mR_{AB}$ is symmetric, so
the transfer matrix of the pair lies in $Sp(4,\mathbb{R})$ and the Wronskian
$(W^{\rm J})^{T}\dot W^{\rm J}-(\dot W^{\rm J})^{T}W^{\rm J}$ is conserved. It vanishes by the
initial data, so $(W^{\rm J})^{T}\dot W^{\rm J}$ is symmetric. The symplectic structure belongs
to the pair and not to the screen matrix itself, which is a $2\times2$ matrix of
$GL(2,\mathbb{R})$ by \eq{SecG22}.

 Consider next equation \eq{SecG9} as a matrix Riccati equation for $\hat{s}_{AB}$ with
$\mR_{AB}$ as input. Substituting the decomposition \eq{SecG7} written for $\hat{s}$, with the
scalars $\theta_{s}$, $\sigma_{s}$ and $\omega_{s}$ of \eq{SecG903}, and using the two
dimensional identities
\beq\label{SecG10}
\sigma_{AC}\sigma_{CB}\,=\,\tfrac12\,\sigma^{2}\,\delta_{AB}\,,\qquad
\omega_{AC}\omega_{CB}\,=\,-\,\tfrac12\,\omega^{2}\,\delta_{AB}\,,\qquad
\sigma_{AC}\omega_{CB}\,+\,\omega_{AC}\sigma_{CB}\,=\,0\,,
\eeq
with 
\beq\label{SecG1001}
\sigma^{2}=\sigma_{AB}\sigma_{AB}\,,\,\,\,\,\omega^{2}=\omega_{AB}\omega_{AB}\,,
\eeq
the quadratic term of \eq{SecG9} is a pure trace and the system splits into
\beqar\label{SecG11}
&\,&\frac{d\theta_{s}}{d\lambda}\,+\,\tfrac12\,\theta_{s}^{2}\,+\,\sigma_{s}^{2}\,-\,
\omega_{s}^{2}\,=\,\mR_{AA}\,,
\\
&\,&\frac{d\sigma_{s\,AB}}{d\lambda}\,+\,\theta_{s}\,\sigma_{s\,AB}\,=\,\mR^{TF}_{AB}\,,
\label{SecG1101}
\\
&\,&\frac{d\omega_{s\,AB}}{d\lambda}\,+\,\theta_{s}\,\omega_{s\,AB}\,=\,\mR_{[AB]}\,=\,0\,,
\label{SecG1102}
\eeqar
where 
\beq\label{SecG1103}
\mR^{TF}_{AB}\,=\,\mR_{AB}-\tfrac12\,\delta_{AB}\,\mR_{CC}\,
\eeq
is its traceless part.
The shear equation carries no
quadratic term because all three products of \eq{SecG10} are multiples of the unit matrix or
vanish. The twist vanishes for $\hat{s}$ as well, $\mR_{AB}$ is symmetric and
$(\hat{s}\hat{s})^{T}=\hat{s}^{T}\hat{s}^{T}=\hat{s}\hat{s}$, so $d\hat{s}/d\lambda$ is
symmetric whenever $\hat{s}$ is and the symmetry propagates from $\hat{s}(-\infty)=0$. Hence
\beq\label{SecG1105}
\omega_{s\,AB}\,\equiv\,0\,,
\eeq
as for $\hat{S}$ in \eq{SecG701}. The same is visible term by term in \eq{SecG902}: that series
is a local iteration, the two factors in $\int^{\lambda}\hat{S}\hat{S}$ standing at the same
$\lambda'$, and the next order is the anticommutator of symmetric matrices at one point, so no
antisymmetric structure can be generated at any order.

 The rotation of the image is a different object then. For symmetric
$A$ and $B$ one has $[A,B]^{T}=-[A,B]$, so the commutator of the generators at two different
points is antisymmetric, and the ordered exponential of symmetric generators is not symmetric.
That antisymmetric part of $W$ is the angle $\Psi$ of \eq{SecG21}, produced by the ordering and
not by any twist.
The two r.h.s. parts in the system of equations have different physical origins. The trace is Ricci,
\beq\label{SecG1104}
\mR_{AA}\,=\,-\,R_{\mu\nu}k^{\mu}k^{\nu}\,=\,-\,\tfrac{\kappa}{2}\,T_{\mu\nu}k^{\mu}k^{\nu}\,,
\eeq
and describes focusing by the matter the line passes through. The traceless part is Weyl and
describes tidal shear, \eq{SecF6} is this statement in light cone language. The
generator $G_{ij}=\tfrac12\D_i\D_j h_{++}$ carries both parts: its trace is
$\tfrac12\D_{\bperp}^{2}h_{++}=R_{++}\propto T_{++}$, which is the Ricci piece and is
nonzero only where the line meets matter, and \eq{SecF9} identifies it with the leading order
current, while it appears in the last structure of \eq{SecD9}. Away from the source the trace
drops and the generator is pure Weyl.

 In our set-up the LO expression for the tidal matrix is provided by \eq{SecD1}, there
$\mR_{AB}=-G_{ij}=-\tfrac12\,\D_i\D_j h_{++}$, the quadratic terms are dropped and
$\lambda\to x^{+}$, so that \eq{SecG702} gives $\hat{S}=\hat{G}$ and \eq{SecG902} makes
$\hat{s}=-\hat{S}$ at this order, and we obtain
\beqar\label{SecG12}
\theta\big|_{LO}\,&=&\,\hat{G}_{ii}\,=\,\tfrac12\,\D_{+}^{-1}\D_{\bperp}^{2}h_{++}\,=\,
-\,\D_i X_{+i}\,,
\\
\sigma_{ij}\big|_{LO}\,&=&\,\hat{G}_{ij}\,-\,\tfrac12\,\delta_{ij}\hat{G}_{kk}\,=\,
-\,\D_i X_{+j}\,+\,\tfrac12\,\delta_{ij}\,\D_k X_{+k}\,,
\label{SecG1201}
\\
\omega\,&=&\,0\,.
\label{SecG1202}
\eeqar
The expansion is then fixed by the source alone and reproduces \eq{SecF9},
\beq\label{SecG13}
\D_{+}\theta\big|_{LO}\,=\,G_{ii}\,=\,R_{++}\,=\,\tfrac{\kappa}{2}\,T_{++}\,,\qquad
\mO_{++}\big|_{LO}\,=\,2\,\D_{+}\hat{S}_{ii}\,=\,2\,\D_{+}\theta\,=\,2\,R_{++}\,.
\eeq
Beyond leading order the two matrices part company,
\beq\label{SecG14}
\hat{S}\,=\,\hat{G}\,-\,\D_{+}^{-1}\mR^{(2)}\,,\qquad
\hat{s}\,=\,-\,\hat{G}\,+\,\D_{+}^{-1}\Le \mR^{(2)}\,-\,\hat{G}\hat{G}\Ra\,+\,\dots\,,
\eeq
$\mR^{(2)}$ being the part of $R_{A+B+}$ quadratic in $h$. The first is a primitive; the second is opposite to it at the first order and carries in addition the primitive of
$\hat{G}\hat{G}$,
which is \eq{SecG902} and which is the term \eq{SecC5} adds to the second order equation.

 The datum of the problem we discuss is the curvature $\mR_{AB}$ and \eq{SecG301} determines
$\hat{S}_{AB}$ from it. With the connection known, \eq{SecG4} is a linear equation which could be resolved by the ordered
exponential of the deformation matrix,
\beq\label{SecG15}
W_{AB}(\lambda)\,=\,\Big[P\exp\int_{-\infty}^{\lambda}\!\!d\lambda'\,\hat{S}(\lambda')\Big]_{AB}\,,
\eeq
the ordering being required in general because $\hat{S}(\lambda_{1})$ and
$\hat{S}(\lambda_{2})$ need not commute at different points of the line; \eq{SecG2102} below shows
the class of profiles for which they do. The ordering can be resolved then on a two dimensional
screen. We obtain a result where the four entries of $W_{AB}$ are replaced by four parameters of the ray
bundle, three of which are unchanged by a rigid rotation of the dyad, $W\to RWR^{T}$, and are
by \eq{SecG23} the only ones the effective current uses. Taking now
the trace part of \eq{SecG7}, we see that it is proportional to the unit matrix and commutes with everything,
so it leaves the ordering and exponentiates by itself. We can write then:
\beq\label{SecG16}
\bW\,=\,a(\lambda)\,\bV\,,\qquad
a\,=\,\exp\Big[\tfrac12\int_{-\infty}^{\lambda}\!\!d\lambda'\,\theta\Big]\,=\,
\sqrt{\det\bW}\,=\,\Delta^{1/2}\,,\qquad
\frac{d\bV}{d\lambda}\,=\,\sigma\,\bV\,,\qquad \det\bV\,=\,1\,.
\eeq
The scalar $a$ is fixed by $\theta$ of \eq{SecG702}, and at the leading order, where
$\hat{s}=-\hat{S}$ and $\theta_{s}=-\theta$ by \eq{SecG903}, it carries the content of the
Raychaudhuri equation of \eq{SecG11}. It is the
square root of the van Vleck and Morette determinant,
normalized by $W(-\infty)=\bone$ so that $\Delta=\det\bW$ with $\Delta(-\infty)=1$, the power being
inverse to the one of the Jacobi matrix by \eq{SecF802}. The
unimodular factor $\bV$ belongs to $SL(2,\mathbb{R})$ and carries the shear alone. Next we recall that
a symmetric traceless matrix on a two dimensional screen has a magnitude and a direction,
\beq\label{SecG17}
\sigma\,=\,s\,\Le \cos2\beta\;\tau_{3}\,+\,\sin2\beta\;\tau_{1}\Ra\,\equiv\,s\;n(\beta)\!\cdot\!\tau\,,
\qquad
\tau_{3}=\begin{pmatrix}1&0\\0&-1\end{pmatrix},\quad
\tau_{1}=\begin{pmatrix}0&1\\1&0\end{pmatrix},\quad
\epsilon=\begin{pmatrix}0&1\\-1&0\end{pmatrix},
\eeq
so that $\pm s$ are the eigenvalues of the shear, $\beta$ is the direction of its principal
axis, and $\sigma_{AB}\sigma_{AB}=2s^{2}$. The same parametrization serves below for the
symmetric factor of $\bV$, with its own angle $\alpha$ of \eq{SecG20} and \eq{SecG22}. The two
angles are different objects. The angle $\beta$ is the direction of the generator at the point,
the angle $\alpha$ is the direction of the distortion accumulated up to that point, and they
coincide only when the ordering is empty. Here $\epsilon_{AB}$ is the antisymmetric symbol,
$\epsilon_{12}=-\epsilon_{21}=1$, the only antisymmetric structure on the screen, so that the
twist of \eq{SecG7}, vanishing there by \eq{SecG701} and nonzero only for the bundle matrix
$\hat{s}$, would be $\omega_{AB}=\tilde\omega\,\epsilon_{AB}$. It satisfies
$\epsilon^{2}=-\bone$, generates rotations
$\exp(\Psi\epsilon)=\cos\Psi\,\bone+\sin\Psi\,\epsilon$ and it anticommutes with
$n\!\cdot\!\tau$. Together with $\bone$ the four matrices span the real $2\times2$ matrices, the
first three symmetric and $\epsilon$ antisymmetric. The product of two shears of different
orientation is a rotation,
\beq\label{SecG18}
[\tau_{3},\tau_{1}]\,=\,2\,\epsilon\,,\qquad
n(\phi_{1})\!\cdot\!\tau\;n(\phi_{2})\!\cdot\!\tau\,=\,\cos2(\phi_{1}-\phi_{2})\,\bone\,+\,
\sin2(\phi_{2}-\phi_{1})\,\epsilon\,,
\eeq
for any two directions $\phi_{1}$ and $\phi_{2}$.
This is a Cartan decomposition, here the symmetric matrices are not a subalgebra, and their
commutators fall into the compact direction $\epsilon$, which spans $\mathfrak{so}(2)$. The
generator $\hat{S}$ lies entirely outside that direction, $\omega_{AB}\equiv0$ by
\eq{SecG701}, so the rotation in $\bW$ is produced by the failure of the shear directions to
close and by nothing in the generator itself. Accordingly \eq{SecG22} below is the polar
decomposition of $GL^{+}(2,\mathbb{R})\cong\mathbb{R}^{+}\times SO(2)\times
SL(2,\mathbb{R})/SO(2)$, with $\Delta^{1/2}$ the scale, $\Psi$ the $SO(2)$ fibre and
$(\Sigma,2\alpha)$ geodesic polar coordinates on the hyperbolic plane, which is why the
hyperbolic functions appear below and not the trigonometric ones.

If the direction of the shear is constant along the line the matrices at different $\lambda$
commute, the ordering in \eq{SecG15} is empty, and
\beq\label{SecG19}
\bV\,=\,\exp\big(\Sigma\,n(\beta)\!\cdot\!\tau\big)\,=\,
\cosh\Sigma\;\bone\,+\,\sinh\Sigma\;n(\beta)\!\cdot\!\tau\,,\qquad
\Sigma\,=\,\int_{-\infty}^{\lambda}\!\!d\lambda'\,s(\lambda')\,.
\eeq
This is the case $\alpha=\beta$, $\Psi=0$.
Then $\bW$ is symmetric, its eigenvalues are $a\,e^{\pm\Sigma}$, and a circular bundle of rays
is mapped into an ellipse of axis ratio $e^{2\Sigma}$ with its axes along $\beta$ and
$\beta+\pi/2$.
If the direction rotates along the line the commutator of \eq{SecG18} generates an $\epsilon$
term. Collecting it by the Magnus expansion,
\beq\label{SecG20}
\bV\,=\,R(\Psi)\,\exp\big(\Sigma\,n(\alpha)\!\cdot\!\tau\big)\,,\qquad
R(\Psi)\,=\,\cos\Psi\;\bone\,+\,\sin\Psi\;\epsilon\,,
\eeq
we have
\beq\label{SecG21}
\Psi\,=\,\int_{-\infty}^{\lambda}\!\!d\lambda_{1}\int_{-\infty}^{\lambda_{1}}\!\!d\lambda_{2}\;
s(\lambda_{1})\,s(\lambda_{2})\;\sin2\big(\beta(\lambda_{2})-\beta(\lambda_{1})\big)\,+\,\dots
\eeq
So, the factorization in \eq{SecG20} is the
polar decomposition of $\bV\in SL(2,\mathbb{R})$ into a rotation and a symmetric unimodular
matrix, and it defines $\Sigma$ and $\alpha$ by
$\bV^{T}\bV=\exp(2\Sigma\,n(\alpha)\!\cdot\!\tau)$. The value of
$\Psi$, given by the second Magnus term, is perturbative in \eq{SecG21}; since $\epsilon$ and $n\!\cdot\!\tau$ anticommute, the
split of $\exp(\Omega_{1}+\Omega_{2})$ into the two factors is fixed by Magnus only up to
$O(h^{3})$. Outside the commuting case of \eq{SecG19} the $\Sigma$ of \eq{SecG20} is the polar
value and not $\int s\,d\lambda$, and the $\alpha$ of \eq{SecG20} is not the $\beta$ of
\eq{SecG17}; the two pairs agree when the ordering does not appear. 
We see then that the angle $\Psi$ starts at order $h^{2}$ and vanishes for a single lens. It is the rotation of
the image of the bundle, $\hat{S}_{AB}$ is
symmetric identically by \eq{SecG301}, and the twist of the congruence is zero by
\eq{SecG1105}, so the rotation is produced by the ordering of shears of different orientation
along the line. This is why $\bW$ is not symmetric although $\hat{S}$ is,
and why $\hat{S}$ has three parameters while $\bW$ has four.

 The criterion for a nontrivial ordering we have in \eq{SecG21} follows from \eq{SecG18}. We
write it for the perturbative generator $\hat{G}$ of \eq{SecC2}, since the expansion at stake
is \eq{SecC7}, and with $s$ and $\beta$ now the magnitude and the direction of its traceless
part; at the leading order these are the $s$ and $\beta$ of \eq{SecG17}. The trace part of
$\hat{G}$ commutes with everything, so
\beq\label{SecG2101}
\big[\hat{G}(x_{1}^{+}),\hat{G}(x_{2}^{+})\big]\,=\,2\,s(x_{1}^{+})\,s(x_{2}^{+})\;
\sin2\big(\beta(x_{2}^{+})-\beta(x_{1}^{+})\big)\;\epsilon\,,
\eeq
and the expansion \eq{SecC7} is an ordered one only when the principal axes of the tidal matrix
rotate along the line.
The commutator represents the ordering, this is seen directly on \eq{SecC7}.
Splitting the square of integration of its second term into the two triangles and relabelling
the integration variables in one of them,
\beq\label{SecG2103}
\int^{x^{+}}\!\!\!dx_{1}^{+}\!\int^{x_{1}^{+}}\!\!\!dx_{2}^{+}\;
\hat{G}(x_{1}^{+})\hat{G}(x_{2}^{+})\,-\,
\tfrac12\Le \int^{x^{+}}\!\!\hat{G}\Ra^{2}\,=\,
\tfrac12\int^{x^{+}}\!\!\!dx_{1}^{+}\!\int^{x_{1}^{+}}\!\!\!dx_{2}^{+}\;
\big[\hat{G}(x_{1}^{+}),\hat{G}(x_{2}^{+})\big]\,,
\eeq
the left hand side being the ordered term minus the unordered product of integrals. Inserting
\eq{SecG2101},
\beq\label{SecG2104}
\tfrac12\int\!\!\int\big[\hat{G}_{1},\hat{G}_{2}\big]\,=\,
\epsilon\int^{x^{+}}\!\!\!dx_{1}^{+}\!\int^{x_{1}^{+}}\!\!\!dx_{2}^{+}\;
s_{1}s_{2}\,\sin2\big(\beta_{2}-\beta_{1}\big)\,=\,\Psi\,\epsilon\,,
\eeq
which is \eq{SecG21}. The image rotation is therefore the ordering itself and it appears at the second order.

 For a factorized profile
\beq\label{SecG2102}
h_{++}\,=\,f(x^{+})\,\Phi(x_{\bperp})\,,\qquad
\hat{G}_{ij}\,=\,\Le \D_{+}^{-1}f\Ra A_{ij}(x_{\bperp})\,,\qquad
\beta\,=\,{\rm const}\,,\qquad \alpha\,=\,\beta\,,\qquad \Psi\,=\,0\,,
\eeq
the ordering is empty and \eq{SecC7} collapses to the exponential of one fixed matrix. This
covers the shock wave of \eq{SecI8}, its smeared version, and any single source with any
longitudinal profile, since $A_{ij}\,=\,\tfrac12\D_i\D_j\Phi$ carries no $x^{+}$ and its
eigendirections at a field point therefore do not move; for a circularly symmetric $\Phi$ they
are the radial and the tangential ones. Elastic scattering
of two point particles at the leading eikonal is of this type. In particular a set of
constituents lying in one plane, $h_{++}=\delta(x^{+})\sum_{n}\Phi_{n}(x_{\bperp}-b_{n})$, is
again of this type: the $A_{n}$ add into a single $A_{\rm tot}$ with one fixed eigenframe and
the line sees one effective lens. What the ordering is sensitive to is the change of the
transverse shape of the field along the line, not its strength, so the constituents have to be
separated in $x^{+}$ as well as in $x_{\bperp}$.
The ordering is nontrivial when
the line crosses sources at different $x^{+}$ which are also separated in $x_{\bperp}$, as in
\eq{SecI17} below; when a Reggeon is emitted between two such crossings, or when the source is
displaced during the crossing. The minimal configuration is two exchanges of different
orientation, which is why $\Psi$ starts at order $h^{2}$.
The same statement is standard in the lensing literature: a single deflector shears an image
but cannot rotate it, while several lens planes can, and the rotation is generated by the lens
to lens coupling at second order \cite{PenMao}. Note that the trivial case is not the same as a
trivial second order term. For the shock the $h^{2}$ current of \eq{SecD6} is nonzero, but the
ordered double integral equals $\tfrac12\big(\int\hat{G}\big)^{2}$ exactly, so the ordering
itself contributes nothing there.

  Now, collecting the factors, \eq{SecG4} could be resolved as
\beq\label{SecG22}
W_{AB}\,=\,\Delta^{1/2}\,
\begin{pmatrix}\cos\Psi & \sin\Psi\\ -\sin\Psi & \cos\Psi\end{pmatrix}
\begin{pmatrix}\cosh\Sigma+\sinh\Sigma\cos2\alpha & \sinh\Sigma\sin2\alpha\\[3pt]
\sinh\Sigma\sin2\alpha & \cosh\Sigma-\sinh\Sigma\cos2\alpha\end{pmatrix}\,,
\eeq
the four parameters being the density of the bundle through $\Delta$, the accumulated shear
$\Sigma$, its orientation $\alpha$ and the image rotation $\Psi$.
The effective current we discussed appears now as follows. Taking the trace of \eq{SecG4} and
using $\epsilon\,n\!\cdot\!\tau=-n\!\cdot\!\tau\,\epsilon$ of \eq{SecG17},
\beq\label{SecG2201}
\tr(\hat{S}\bW)\,=\,d(\tr\bW)/d\lambda\,,\,\,\,\,\tr[R(\Psi)\,n\!\cdot\!\tau]\,=\,0\,,
\eeq
the second relation making the orientation $\alpha$ drop out of the trace. With $c=2$ of \eq{SecC6},
\beq\label{SecG23}
\tr\bW\,=\,2\,\Delta^{1/2}\,\cos\Psi\,\cosh\Sigma\,,\qquad
\mO_{++}\,=\,2\,\frac{d^{2}}{d\lambda_{+}^{2}}\,\tr\bW\,=\,
4\,\frac{d^{2}}{d\lambda_{+}^{2}}\Big[\Delta^{1/2}\,\cos\Psi\,\cosh\Sigma\Big]\,.
\eeq
The effective current is built from three of the four parameters and does not see the direction
of the shear; in the shear free case $\Sigma\,=\,\Psi\,=\,0$ it reduces to
$\mO_{++}=4\,d^{2}\Delta^{1/2}/d\lambda_{+}^{2}$, the second light cone derivative of the
square root of the van Vleck determinant alone, which by \eq{SecG29} is the prefactor of the eikonal
amplitude.

 As we underlined already, at leading order the whole matrix is the transverse Hessian of one scalar. Writing
\beq\label{SecG24}
\varphi\,=\,-\,\tfrac12\,\D_{+}^{-2}h_{++}\,,\qquad
X_{+i}\,=\,\D_i\D_{+}\varphi\,,\qquad
\hat{G}_{ij}\,=\,-\,\D_i\D_j\D_{+}\varphi\,,
\eeq
\eq{SecG15} truncated at the first term, together with \eq{SecG12} and \eq{SecG22}, gives
\beq\label{SecG25}
W_{ij}\big|_{LO}\,=\,\delta_{ij}\,-\,\D_i\D_j\varphi\,,\qquad
W^{\rm J}_{ij}\big|_{LO}\,=\,\delta_{ij}\,+\,\D_i\D_j\varphi\,,\qquad
\Delta^{1/2}\big|_{LO}\,=\,1\,-\,\tfrac12\,\D_{\bperp}^{2}\varphi\,+\,O(h^{2})\,,
\eeq
\beq\label{SecG2500}
\Sigma\big|_{LO}\,=\,\Big[\tfrac12\,\sigma^{(1)}_{ij}\sigma^{(1)}_{ij}\Big]^{1/2}\,+\,O(h^{2})\,,
\qquad
\Psi\big|_{LO}\,=\,0\,,
\eeq
with $\sigma^{(1)}_{ij}$ the traceless part of $W_{ij}\big|_{LO}-\delta_{ij}$,
\beq\label{SecG2501}
\sigma^{(1)}_{ij}\,=\,-\,\D_i\D_j\varphi\,+\,\tfrac12\,\delta_{ij}\D_{\bperp}^{2}\varphi\,.
\eeq
The matrix $W^{\rm J}_{ij}$ is the Jacobi matrix of gravitational lensing and $W_{ij}$ is its
inverse, the amplification matrix; $\varphi$ is the deflection potential of the line.
The expansion of \eq{SecC7} is the expansion of \eq{SecG22} around $\Delta^{1/2}=1$,
$\Sigma=\Psi=0$.

  The question we can ask is how some scattering data can be obtained directly in the framework of
the formalism. What follows are the classical observables of test rays crossing the field of
the other line.
Taking again the leading order component $h_{++}$ above flat space time, the
only Christoffel symbol entering the transverse motion is
$\Gamma^{i}_{++}=\tfrac12\D_i h_{++}$, so with $x^{+}$ as the parameter the null geodesic obeys
\beq\label{SecG2502}
\frac{d^{2}x_{i}}{d(x^{+})^{2}}\,=\,-\,\tfrac12\,\D_i h_{++}\big(x^{+},x(x^{+})\big)\,,
\eeq
or
\beq\label{SecG2600}
\frac{dx_{i}}{dx^{+}}\,=\,\int^{x^{+}}\!\!dx_{1}^{+}
\Le -\tfrac12\D_i h_{++}\Ra\!\big(x_{1}^{+},x(x_{1}^{+})\big)\,\simeq\,
X_{+i}\big(x^{+},x(x^{+})\big)\,
\eeq
with
\beq\label{SecG2601}
x_{i}(-\infty)\,=\,b_{i}\,,\qquad\,J_{ij}\,=\,\frac{\partial x_{i}(x^{+})}{\partial b_{j}}\,.
\eeq
In the last step the integrand stands at the position the ray had at
$x_{1}^{+}$, while $X_{+i}$ of \eq{SecB3} carries it at the current position, and the two
agree at the first order in $h$.
The chain rule and $\hat{G}_{ij}=-\D_i X_{+j}$ of \eq{SecD2} give
\beq\label{SecG2602}
\frac{dJ_{ij}}{dx^{+}}\,=\,\big(\D_k X_{+i}\big)\,J_{kj}\,=\,-\,\hat{G}_{ki}\,J_{kj}\,=\,
-\,\hat{G}_{ik}\,J_{kj}\,,\qquad J_{ij}(-\infty)\,=\,\delta_{ij}\,,
\eeq
the last step using that $\hat{G}_{ij}=\tfrac12\D_{+}^{-1}\D_i\D_j h_{++}$ is symmetric.
This is \eq{SecC4} with the opposite generator, so $J_{ij}$ is the Jacobi matrix of
\eq{SecF801} at this order and not $W_{ij}$, the two being related by \eq{SecF802},
\beq\label{SecG26}
W^{\rm J}_{ij}\,=\,\frac{\partial x^{\rm out}_{i}}{\partial b_{j}}\,,\qquad
\bW\,=\,\Big[\big(\bW^{\rm J}\big)^{T}\Big]^{-1}\,+\,O(h^{2})\,,\qquad
\D_{+}W^{\rm J}_{ij}\big|_{+\infty}\,=\,\frac{\partial \Theta_{i}}{\partial b_{j}}\,,\qquad
\Theta_{i}\,=\,X_{+i}(+\infty)\,,
\eeq
with $b_{i}$ the impact parameter of the ray and $\Theta_{i}$ its deflection, the second
relation being \eq{SecG2602} before the integration. Here $x^{\rm out}$ is the trajectory of
the field $X_{+i}$, which is the null geodesic at first order in $h$; the exact geodesic map is
$\bW^{\rm J}$ of \eq{SecI10}, and the two differ as \eq{SecI11} describes. The four
parameters of \eq{SecG22} are therefore the four parameters of a narrow bundle around a chosen
ray. For an incident flux of rays uniform in $b$, with $N$ their number, the densities on the
screen and in the deflection angle are
\beq\label{SecG27}
\frac{dN}{d^{2}x^{\rm out}}\,=\,\big|\det\bW^{\rm J}\big|^{-1}\,=\,\det\bW\,=\,\Delta\,,\qquad
\frac{dN}{d^{2}\Theta}\,=\,\big|\det\D_{+}\bW^{\rm J}(+\infty)\big|^{-1}\,=\,
\frac{1}{\big|\det\hat{s}(+\infty)\big|\;\det\bW^{\rm J}(+\infty)}\,.
\eeq
The positional density is $\Delta$ itself, fixed by the scale parameter $\Delta^{1/2}$ alone, while the angular one needs the
connection at the end of the line as well. The principal magnifications are the singular
values of the Jacobi matrix, the singular values of \eq{SecG22} being their inverses,
\beq\label{SecG28}
\Lambda_{1,2}\,=\,\Delta^{-1/2}\,e^{\pm\Sigma}\,,\qquad
\bW^{T}\bW\,=\,\Delta\,\exp\big(2\Sigma\,n(\alpha)\!\cdot\!\tau\big)\,,\qquad
\det\bW\,=\,\Delta\,,
\eeq
the $\Lambda_{1,2}$ being the singular values of $\bW^{\rm J}$ and those of $\bW$ their
inverses, the rotation dropping out of both.
They are the eigenvalues of $\bW^{\rm J}$ only at $\Psi=0$: for $\Psi\neq0$ one has
$\tr\bW=2\Delta^{1/2}\cos\Psi\cosh\Sigma$, so the eigenvalues are
complex whenever $|\cos\Psi\cosh\Sigma|<1$, while the magnifications stay real and positive.
Here $\Sigma$ gives the axis ratio $e^{2\Sigma}$ of the image, $\alpha$ its orientation and
$\Psi$ its rotation. A caustic is a zero of one of them,
\beq\label{SecG2801}
\Lambda_{1}\Lambda_{2}\,=\,\det\bW^{\rm J}\,=\,\Delta^{-1}\,,
\eeq
so caustic is defined through
\beq\label{SecG2802}
\det\bW^{\rm J}\,=\,0\,,
\eeq
one eigenvalue vanishing while the other stays finite, which by \eq{SecG28} requires
$\Delta^{-1/2}\to0$ and $\Sigma\to\infty$ together, so that the parametrization \eq{SecG22}
degenerates there.
The condition is written for the Jacobi matrix $\bW^{\rm J}$ of \eq{SecF801} and not for $\bW$.
The reason is \eq{SecG16}: $\det\bW=\exp\int_{-\infty}^{\lambda}d\lambda'\,\theta$ with $\theta=\tr\hat{S}$
finite, so $\det\bW$ is a real exponential and is strictly positive for any profile and has no zero at any
$\lambda$. The matrix $\bW$ is an ordered exponential and cannot leave $GL^{+}(2,\mathbb{R})$, while
$\bW^{\rm J}$ obeys the second order equation \eq{SecG9} and does degenerate. The shock of \eq{SecI14}
is the explicit case: $\det\bW=1$ and no caustic, against $\det\bW^{\rm J}=1-(ax^{+})^{2}$ of \eq{SecI13}
which vanishes at the Einstein ring \eq{SecI1501}. The two are inverse at the first order in
$h$, $\hat{s}=-\hat{S}$ by \eq{SecG902}, which is the accuracy of the caustic statement.
The caustic then fixes the critical impact parameter. Since $\Delta^{1/2}$ is the
prefactor of the eikonal propagator and each caustic advances the Morse index,
\beq\label{SecG29}
\mA\,\sim\,\Delta^{1/2}\,e^{\,i\chi\,-\,i\pi n_{c}/2}\,,\qquad
n_{c}\,=\!\!\sum_{\det\bW^{\rm J}=0}\!\!\dim\ker\bW^{\rm J}\,,
\eeq
where $n_{c}$ is the Morse index of the segment, the caustics between $-\infty$ and the
observation point counted with multiplicity, which is one or two on a two dimensional screen.
It carries the branch of the square root: $\det\bW^{\rm J}$ changes sign at a simple zero and
$\Delta^{1/2}$ continues through it with an extra $e^{-i\pi/2}$. It refers to $\bW^{\rm J}$,
while the $\Delta$ of \eq{SecG16} is built on $\bW$; by \eq{SecF802} the two are the same $\Delta$ at
the leading order, $\det\bW=1/\det\bW^{\rm J}$, which is the accuracy of the present statement. For
the shock \eq{SecI14} gives $\det\bW=1$ for the perturbative matrix and no caustic at all. The
determinant supplies both the one loop weight of the saddle point and the phase it collects.

 As a demonstrating example we again take the matrix $W^{\rm J}_{ij}=\delta_{ij}+\D_i\D_j\varphi$ of \eq{SecG25} with the
potential $\varphi$ of \eq{SecG24}. Splitting the Hessian into its trace and traceless parts in
the basis of \eq{SecG17},
\beq\label{SecG2901}
\D_i\D_j\varphi\,=\,\tfrac12\,\delta_{ij}\D_{\bperp}^{2}\varphi+\gamma_{1}\tau_{3}+\gamma_{2}\tau_{1}
\eeq
with 
\beq\label{SecG2902}
\gamma_{1}\,=\,\tfrac12(\D_{1}^{2}-\D_{2}^{2})\varphi\,,\,\,\,\,\gamma_{2}\,=\,\D_{1}\D_{2}\varphi
\eeq
and using 
\beq\label{SecG2903}
(\gamma_{1}\tau_{3}+\gamma_{2}\tau_{1})^{2}\,=\,|\gamma|^{2}\bone
\eeq
its eigenvalues are
\beq\label{SecG30}
\Lambda_{1,2}\,=\,1\,+\,\tfrac12\,\D_{\bperp}^{2}\varphi\,\pm\,|\gamma|\,,\qquad
|\gamma|\,=\,\sqrt{\gamma_{1}^{2}+\gamma_{2}^{2}}\,,
\eeq
and the parameters of \eq{SecG22} follow by inverting \eq{SecG28},
\beq\label{SecG3001}
\Delta^{-1/2}\,=\,\sqrt{\Lambda_{1}\Lambda_{2}}\,,\qquad
\Sigma\,=\,\tfrac12\,\ln\frac{\Lambda_{1}}{\Lambda_{2}}\,,\qquad
\alpha\,=\,\alpha_{\gamma}\,+\,\tfrac{\pi}{2}\,,\qquad
\tan2\alpha_{\gamma}\,=\,\frac{\gamma_{2}}{\gamma_{1}}\,,\qquad \Psi\,=\,0\,.
\eeq
The tangent alone fixes $2\alpha$ modulo $\pi$. The branch is settled by the traceless part of
$\bW$ being $-\gamma\!\cdot\!\tau$, so that $(\cos2\alpha,\sin2\alpha)$ is antiparallel to
$(\gamma_{1},\gamma_{2})$: the inverse matrix exchanges the two principal axes.
We underline that \eq{SecG25} and \eq{SecG3001} describe the same matrix
$\delta_{ij}+\D_i\D_j\varphi$ and are not two answers for one quantity. \eq{SecG3001} is the
precise inversion of \eq{SecG28} and uses no expansion, so it is exact for that matrix, while
\eq{SecG25} is its leading order truncation,
\beq\label{SecG3003}
\Delta^{-1/2}\,=\,1\,+\,\tfrac12\,\D_{\bperp}^{2}\varphi\,-\,\tfrac12\,|\gamma|^{2}\,+\,\dots\,,
\qquad
\Sigma\,=\,|\gamma|\,-\,\tfrac12\,|\gamma|\,\D_{\bperp}^{2}\varphi\,+\,\dots\,,
\eeq
the second structures being the $O(h^{2})$ terms dropped there. They cancel against the
expansion of the exponential in \eq{SecG28}, which returns \eq{SecG30}, so the truncated forms
must not be inserted into \eq{SecG2801}. Keeping the exact forms is meaningful where the matrix
itself is exact, which is the thin lens of \eq{SecI10}, or where $\D_i\D_j\varphi$ is resummed
while other second order terms are dropped, which is the eikonal approximation.
So, the caustic condition itself is
\beq\label{SecG3002}
\det\bW^{\rm J}\,=\,\Lambda_{1}\Lambda_{2}\,=\,0
\quad\rightarrow\quad
\tfrac12\,\D_{\bperp}^{2}\varphi\,\pm\,|\gamma|\,=\,-\,1\,,
\eeq
the critical curve of the lens; away from the matter $\D_{\bperp}^{2}\varphi=0$ and it reduces
to $|\gamma|=1$. It requires the entries of $\D_i\D_j\varphi$ to be of order one, where the
linearized forms fail while \eq{SecG30} and \eq{SecG3001} stay correct. For a thin lens they
are exact, since $\bW^{\rm J}=\bone-x^{+}A$ of \eq{SecI10} holds to all orders, and they return
\eq{SecI13}, see discussion further.
The deflection and the eikonal phase could be calculated as follows then:
\beq\label{SecG3004}
\D_{j}\Theta_{i}=\Le \hat{s}\bW^{\rm J}\Ra_{ij}(+\infty)\,=\,-\,\hat{S}_{ij}(+\infty)\,+\,O(h^{2})
\eeq
by \eq{SecG26}, and $W^{\rm J}_{ij}-\delta_{ij}=\D_i\D_j\varphi$, so
\eq{SecG22} fixes $\Theta_{i}$ up to a constant and $\chi$ up to a function linear in $b$, both
fixed by $\Theta_{i}\to0$ at large $b$.

\section{Covariant interaction and eikonal Wilson lines}\label{SecH}

 Having a covariant description of the bundle of rays propagating in an external field, we return to our problem of high energy gravitational scattering.
The problem is dynamical, so the covariant form of the interaction which describes the scattering through the
Reggeized fields is 
\beq\label{SecH1}
L_{int}\,=\,\frac{1}{\kappa}\;\mB_{\bar k\bar k}\;
\frac{D}{d\lambda_{+}}\Big[\hat{S}^{(+)}_{AB}\,W^{(+)}_{BA}\Big]\;,
\eeq
where
\beq\label{SecH2}
\mB_{\bar k\bar k}\,=\,\mB_{\mu\nu}\bar k^{\mu}\bar k^{\nu}\,,\,\,\,\,
\mB_{kk}\,=\,\mB_{\mu\nu}k^{\mu}k^{\nu}\,. 
\eeq
In turn, the current to current interaction \eq{SecF14} is
\beq\label{SecH3}
L_{\rm int}\,=\,\frac{1}{\kappa}\;
\Big(\frac{D}{d\lambda_{+}}\Big[\hat{S}^{(+)}_{AB}\,W^{(+)}_{BA}\Big]\Big)\;\Box_{S}^{-1}\;
\Big(\frac{D}{d\lambda_{-}}\Big[\hat{S}^{(-)}_{CD}\,W^{(-)}_{DC}\Big]\Big)\,,
\eeq
where $\Box_{S}$ is the Laplacian on the screen, the two dimensional surface orthogonal to
$k$ and $\bar k$, which reduces to $\D_{\bperp}^{2}$ in the light cone frame. With
$\mR^{(+)}_{AA}\,=\,-R_{\mu\nu}k^{\mu}k^{\nu}$ of \eq{SecG1104}, which is exact, and with
$W^{(\pm)}_{AB}\to\delta_{AB}$ at the leading order, \eq{SecH3} becomes
\beq\label{SecH4}
L_{\rm int}\,=\,
\tfrac{\kappa}{4}\,T_{\mu\nu}k^{\mu}k^{\nu}\,\Box_{S}^{-1}\,T_{\rho\sigma}\bar k^{\rho}\bar k^{\sigma}\,.
\eeq
Together with \eq{SecE8} and \eq{SecF14} this is the third of three equivalent writings of the
same interaction: in terms of the operators $\mO_{\pm\pm}$, in terms of the Riemann tensor, and
covariantly on the congruence. They differ only in the variables, not in content. As we see then, in all these formulations a new ordered object appears which is different from the regular Wilson lines built on the basis of the graviton field, see for examples \cite{Alessio,NS,White,White1,MNSW,GWL1,GWL3,GWL2}. The corresponding induced effective currents, which
appear in the equations of motion, are derived in Appendix~\ref{AppB}.

 Proceeding, we notice an additional and very interesting variant of the effective action form: we could write \eq{SecH3} in the parameters of \eq{SecG22} and 
then we obtain the following. The transport operator is generated by $\hat{S}$:
\eq{SecG4},
\beq\label{SecH401}
\frac{D\bW}{d\lambda}\,=\,\hat{S}\,\bW\,,\qquad \bW\big|_{\lambda\to-\infty}\,=\,\bone\,,
\eeq
so the trace of \eq{SecH401} gives 
\beq\label{SecH402}
\hat{S}_{AB}\,W_{BA}\,=\,\tr\Le \hat{S}\,\bW\Ra\,=\,\tr\frac{D\bW}{d\lambda}\,=\,
\frac{d}{d\lambda}\,\tr\bW\,,
\eeq
which is the first relation of \eq{SecG2201}. By \eq{SecG16} and \eq{SecG20} we have that
\beq\label{SecH40201}
\bW\,=\,a\,R(\Psi)\,M\,,\,\,\, a\,=\,\Delta^{1/2}
\eeq
and
\beq\label{SecH403}
M\,=\,\exp\big(\Sigma\,n(\alpha)\!\cdot\!\tau\big)\,=\,
\cosh\Sigma\;\bone\,+\,\sinh\Sigma\;n(\alpha)\!\cdot\!\tau\,,\qquad
R(\Psi)\,=\,\cos\Psi\;\bone\,+\,\sin\Psi\;\epsilon\,.
\eeq
The matrices $\tau_{3}$, $\tau_{1}$ and $\epsilon$ of \eq{SecG17} are traceless, and so is
\beq\label{SecH404}
\epsilon\,n(\alpha)\!\cdot\!\tau\,=\,
\cos2\alpha\;\epsilon\tau_{3}\,+\,\sin2\alpha\;\epsilon\tau_{1}\,=\,
-\,\cos2\alpha\;\tau_{1}\,+\,\sin2\alpha\;\tau_{3}\,,
\eeq
and we obtain for the trace
\beq\label{SecH405}
\tr\bW\,=\,2\,a\,\cos\Psi\,\cosh\Sigma\,=\,2\,\Delta^{1/2}\,\cos\Psi\,\cosh\Sigma\,,
\eeq
which is \eq{SecG23} and in which $\alpha$ is absent. Putting \eq{SecH405} into \eq{SecH402}
and differentiating once more along the ray,
\beq\label{SecH406}
\frac{D}{d\lambda_{\pm}}\Big[\hat{S}^{(\pm)}_{AB}\,W^{(\pm)}_{BA}\Big]\,=\,
\frac{d^{2}}{d\lambda_{\pm}^{2}}\,\tr\bW^{(\pm)}\,=\,
2\,\frac{d^{2}}{d\lambda_{\pm}^{2}}
\Big[\Delta_{\pm}^{1/2}\,\cos\Psi_{\pm}\,\cosh\Sigma_{\pm}\Big]\,,
\eeq
so that \eq{SecH3} becomes
\beq\label{SecH407}
L_{\rm int}\,=\,\frac{4}{\kappa}\,
\Le \frac{d^{2}}{d\lambda_{+}^{2}}
\Big[\Delta_{+}^{1/2}\cos\Psi_{+}\cosh\Sigma_{+}\Big]\Ra\;\Box_{S}^{-1}\;
\Le \frac{d^{2}}{d\lambda_{-}^{2}}
\Big[\Delta_{-}^{1/2}\cos\Psi_{-}\cosh\Sigma_{-}\Big]\Ra\,.
\eeq
The rewriting is exact and no matrix and no ordering symbol is left in it. The interaction is a
bilinear form in six scalar functions on the screen, three on each line: the area of the ray
bundle through $\Delta^{1/2}$, the accumulated shear $\Sigma$ and the image rotation $\Psi$.
The orientation $\alpha$ of the shear drops out by the second relation of \eq{SecG2201}.
We notice that the $\Sigma$ and the $\Psi$ standing there are defined by the polar decomposition \eq{SecG20}
of the ordered exponential itself and not by the generator at a point, so obtaining them from
$\hat{S}$ is the ordering problem again. 

 We return now to the comparison between different types of operators. In order to understand the difference, consider the object used in the eikonal literature, see again \cite{Alessio,NS,White,White1,MNSW,GWL1,GWL3,GWL2}, which is
\beq\label{SecH5}
\mathcal{W}_{\rm eik}\,=\,\exp\Big(\tfrac{i}{2}\,p^{\mu}p^{\nu}\!\int\! ds\;h_{\mu\nu}\Big)\;\rightarrow\;e^{i\chi}\,,\qquad
\chi\,=\,\tfrac{1}{2}\,p^{+}\!\int dx^{+}h_{++}\,,
\eeq
where on the plus line $p^{\mu}p^{\nu}h_{\mu\nu}\to p^{+} p^{+} h_{++}$ and $ds=dx^{+}/p^{+}$.
The two constructions share the same null line, the exponentiation, the Born term $\tfrac{\kappa}{4}\,s^{2}/t$ and appear in the same physical
processes. But, first of all, the two forms of the exponential differ as an Abelian one against a
matrix one. The $\mathcal{W}_{\rm eik}$ is a $U(1)$ phase, its exponent is a number and
the ordering does nothing. $W_{ij}$ is a $GL(2,\mathbb R)$ matrix on the transverse plane, the pair
$(W,\D_{+}W)$ of its second order form \eq{SecG401} being $Sp(4,\mathbb R)$, the generator
$-\mR+\hat{S}\hat{S}$ there being symmetric, and the ordering is
non-commutative. This is the structure similar to the one supplied
in QCD by the color algebra and absent from the Abelian gravitational line. The second issue is about
which transverse moment is resummed in the operators. The Abelian $\mathcal{W}_{\rm eik}$ resums the zeroth moment
$\int dx^{+}h_{++}$, the phase. $W_{ij}$ resums the second moment
$\D_{+}^{-2}\D_i\D_j h_{++}$ which is minus the twice integrated tidal matrix $\mE^{(K)}_{ij}$ of
\eq{SecC11}. Both provide a leading contribution to the quantities under consideration, because the extra
$\int dx^{+}\D_{\bperp}^{2}$ per insertion is $O(1)$ in the Glauber region, which is the
statement that $h_{++}$ has negative scaling dimension \cite{BK}, see \cite{BK1} for the
counting beyond the leading power. Next we account for the gauge status
of the exponentials. The exponent of \eq{SecH5} is not diffeomorphism invariant and needs the
cusp regularization. The \eq{SecF7} exponent, in turn, is invariant at linear order, so
\eq{SecH3} consists of two invariant transports. It is interesting to remark also about
what each object carries. By the \eq{SecG23} form we see that the current is built from $\Delta$, $\Sigma$ and
$\Psi$ and not from $\alpha$, so the orientation of the distortion enters the action only
through the transverse profiles of the other three. Of the three, $\Psi$ is the one with no
counterpart in the Abelian phase: it is of order $h^{2}$ and needs two exchanges of different
orientation. Conversely $W_{ij}$ carries no longitudinal information,
it resums the transverse Hessian of the phase and not the phase itself, so it fixes $\chi$ only
up to a function linear in $b$. The $b$ independent part of $\chi$
is the shift in $x^{-}$ across the line, $\Delta x^{-}\,=\,-\chi/p^{+}$ for the shock of
\eq{SecI8} further, which is the Shapiro delay and which the exponent of \eq{SecH5} does carry. The two
objects are complementary in that sense.

 There is another observation then, that the two objects can be written as one because the phase generator is a multiple of the unit matrix. With
$\dot\chi=\tfrac{1}{2}p^{+}h_{++}$ from \eq{SecH5}, the matrix $\dot\chi\,\delta_{ik}$ commutes
with $\hat{S}_{ik}$, leaves the ordering and exponentiates by itself, exactly as the trace part
does in \eq{SecG16}. Therefore
\beq\label{SecH1001}
\bU_{ik}\,=\,\Big[P\exp\int dx^{+}\Le \,\tfrac{i}{2}\,p^{+}h_{++}\,\delta_{ik}\,+\,
\hat{S}_{ik}\,\Ra\Big]\,=\,e^{\,i\chi}\,W_{ik}\,,
\eeq
a single ordered exponential along the line. The Abelian $\mathcal{W}_{\rm eik}$ and the matrix
$W_{ik}$ are the $U(1)$ and the $GL(2,\mathbb{R})$ factors of one $GL(2,\mathbb{C})$ transport,
whose generator is a real symmetric matrix plus an imaginary multiple of the unit one, with a
complex expansion
\beq\label{SecH1002}
\tr\Le \hat{S}\,+\,i\,\dot\chi\,\bone\Ra\,=\,\theta\,+\,2\,i\,\dot\chi\,,
\eeq
the real part being the trace $\theta$ of \eq{SecG8}, minus the focusing, and the imaginary part
the rate of the phase.
The object \eq{SecH1001} composes correctly along the line, the phases adding and the matrices
ordering, so it is a transfer matrix. Namely, it is not a regular Green function, since it carries no free
transverse spreading and reduces to $\delta_{ik}$ and not to the free propagator, it is a kind of truncated 
propagator of the particle in an external field, see \cite{OurZubkov1,OurZubkov2,Meggio,Meggio1,Meggio2,Meggio3,Meggio4,Meggio5}.
In an amplitude the same combination appears through the saddle point.
The saddle point is taken in the impact parameter. Writing
\beq\label{SecH100201}
K(x_{f})\,=\,\int\! d^{2}b\;A(b)\;
e^{\,i\,\Le S_{\rm free}(x_{f},b)\,+\,\chi(b)\Ra}\,,\qquad
S_{\rm free}\,=\,-\,\frac{p^{+}}{2x^{+}}\,\big(x_{f}-b\big)^{2}\,,
\eeq
the sign of $S_{\rm free}$ following from $\eta_{ij}=-\delta_{ij}$ of \eq{SecB9}. Stationarity of \eq{SecH100201} in $b$,
$\D_{b_{i}}\Le S_{\rm free}+\chi\Ra=0$, gives
$x_{f\,i}\,=\,b_{i}\,-\,(x^{+}/p^{+})\,\D_{i}\chi\,=\,x^{\rm out}_{i}(b)$, which is the deflected ray of
\eq{SecG26}, and the fluctuation matrix of the Gaussian integration is
\beq\label{SecH1003}
\frac{\partial^{2}S}{\partial b_{i}\,\partial b_{j}}\,=\,
-\,\frac{p^{+}}{x^{+}}\,\delta_{ij}\,+\,\D_i\D_j\chi\,=\,
-\,\frac{p^{+}}{x^{+}}\,\Le \delta_{ij}\,-\,\frac{x^{+}}{p^{+}}\,\D_i\D_j\chi\Ra\,.
\eeq
The field independent factor assembles the free propagator $K_{0}$, and by \eq{SecH12}
\beq\label{SecH100301}
\delta_{ij}\,-\,\frac{x^{+}}{p^{+}}\,\D_i\D_j\chi\,=\,\delta_{ij}\,-\,x^{+}A_{ij}\,=\,
W^{\rm J}_{ij}\,+\,O(h^{2})\,,
\eeq
so the ratio of the two prefactors is $(\det\bW^{\rm J})^{-1/2}=(\det\bW)^{1/2}=\Delta^{1/2}$. The
same factor follows from \eq{SecG27} with no saddle point at all, there the ray density on the screen
is $|\det\bW^{\rm J}|^{-1}=\det\bW$ and the modulus of the wave is its square root. We underline that the determinant comes from the Hessian in $b$ and not from the one
in $x_{f}$.
Writing $\xi_i=x_{f\,i}-x^{\rm out}_i(b)$ for the deviation from the trajectory which starts at
the impact parameter $b$, the Gaussian saddle point form of the remainder follows from \eq{SecH100301},
\beq\label{SecH10}
\frac{K(x_{f},b)}{K_{0}(x_{f},b)}\,\propto\,\big(\det \bW^{\rm J}\big)^{-1/2}\,
\exp\Big(\,i\chi(b)\,+\,i\,\xi_i\,\D_i\chi(b)\,+\,\frac{i}{2}\,\xi_i\,\big(\D_i\D_j\chi\big)\,\xi_j\,\Big)\,,
\eeq
the phase on the ray, the curvature of that phase transverse to it, and the determinant as the
normalization, all in one expression. At $h=0$ the ratio is unity, and its first correction
$1+i\chi+\dots$ is a pure phase, which is the statement that the leading eikonal is elastic and
changes no modulus. Both structures are terms of a single Taylor expansion of
the eikonal phase about the trajectory,
\beq\label{SecH11}
\chi(x_{f})\,=\,\underbrace{\chi(b)}_{\mathcal{W}_{\rm eik}}\,+\,
\underbrace{\xi_i\,\D_i\chi(b)}_{-\,p^{+}\Theta_{i}}\,+\,
\underbrace{\tfrac12\,\xi_i\,\big(\D_i\D_j\chi\big)\,\xi_j}_{W_{ij}}\,+\,\dots\,,
\eeq
the linear term being the deflection, $\D_i\chi\,=\,-\,p^{+}\Theta_{i}$ with $\Theta_{i}$ of
\eq{SecG26}, since the stationarity of \eq{SecH100201} holds in $b$ and not in $\xi$. The Hessian
here is the generator of $W$ up to a constant: by \eq{SecH9} and \eq{SecG301}, with the retarded
primitive of \eq{SecH702},
\beq\label{SecH12}
\D_i\D_j\chi\,=\,p^{+}\,A_{ij}\,=\,p^{+}\,\hat{S}_{ij}\,,\qquad x^{+}>0\,,
\eeq
so \eq{SecH11} is the statement about the zeroth, the first and the second transverse moment made
above, written in one line. On the ray itself only $\chi(b)$ survives, and the deflection is its
gradient in the impact parameter; that gradient is all the Abelian line carries.

 Finally, as an example, consider a shock graviton's profile 
\beq\label{SecH6}
h_{++}=\delta(x^{+})\Phi(x_{\bperp}).
\eeq
Then 
\beq\label{SecH7}
\chi=p^{+}\,\Phi / 2\,.
\eeq
Using \eq{SecG24} result we have in turn, with $\vartheta(x^{+})$ the step function,
$\vartheta=1$ for $x^{+}>0$ and $\vartheta=0$ otherwise, written so as not to collide with the
expansion $\theta$ of \eq{SecG7} or with the deflection $\Theta_{i}$ of \eq{SecG26},
\beq\label{SecH701}
\D_{+}^{-1}h_{++}\,=\,\vartheta(x^{+})\,\Phi\,,\qquad
\D_{+}^{-2}h_{++}\,=\,x^{+}\vartheta(x^{+})\,\Phi\,,
\eeq
so that, using \eq{SecG24} and then \eq{SecH7},
\beq\label{SecH702}
\varphi\,=\,-\,\tfrac12\,\D_{+}^{-2}h_{++}\,=\,-\,\tfrac12\,x^{+}\vartheta(x^{+})\,\Phi\,=\,
-\,\frac{x^{+}\vartheta(x^{+})}{p^{+}}\,\chi(x_{\bperp})\,.
\eeq
The retarded primitive is used here; with the symmetric $\D_{+}^{-1}$ of \eq{SecC8} the second
primitive of a $\delta$ profile does not exist without the regularization of \eq{SecE10}. With
this, for $x^{+}>0$,
\beq\label{SecH8}
\mE^{(K)}_{ij}\,=\,\tfrac12\,\D_{+}^{-2}\D_i\D_j h_{++}\,=\,\frac{x^{+}}{p^{+}}\,\D_i\D_j\chi\,=\,-\,\D_i\D_j\varphi,
\qquad
W_{ij}\,=\,\big[\exp\big(x^{+}A\big)\big]_{ij}\,,\qquad
A_{ij}\,=\,\frac{1}{p^{+}}\,\D_i\D_j\chi\,.
\eeq
In the expression $A_{ij}$ carries no $x^{+}$, so the generator commutes with itself along the line and the ordering of \eq{SecC3} is 
absent. Expanding it to the first order returns \eq{SecG25}, and the thin lens Jacobi matrix of
\eq{SecI10} is its inverse transposed by \eq{SecF802}, itself exact for the shock. Therefore
\beq\label{SecH9}
\mathcal{W}_{\rm eik}\,=\,\exp\big(i\chi\big)\,,\qquad
W_{ij}\,=\,\big[\exp\big(x^{+}A\big)\big]_{ij}\,,\qquad
A_{ij}\,=\,\frac{1}{p^{+}}\,\D_i\D_j\chi\,.
\eeq
Concluding, we see that the two objects are the phase and the Hessian of the same eikonal. In WKB language the eikonal
propagator is $\Delta^{1/2}e^{i\chi}$ with $\Delta$ the van Vleck and Morette determinant,
$\Delta=1/\det\bW^{\rm J}=\det\bW$ at the leading order by \eq{SecF802}.

\section{Working example: shock wave profile}\label{SecI}

  The shock wave is the one background on which both the perturbative generator of
\eq{SecC2} and the geometric generator of \eq{SecG9} can be integrated in closed form, so it
fixes the precise sense in which the two transports differ. The null congruence, its
optical scalars and the van Vleck determinant in this background are treated in \cite{Shore},
where the same Jacobi matrix appears through the Penrose limit. Take
\beq\label{SecI8}
h_{++}\,=\,\delta(x^{+})\,\Phi(x_{\bperp})\,,\qquad
A_{ij}\,=\,\tfrac12\,\D_i\D_j\Phi\,,\qquad
G_{ij}\,=\,\delta(x^{+})\,A_{ij}\,,\qquad
\hat{G}_{ij}\,=\,\vartheta(x^{+})\,A_{ij}\,,
\eeq
with $X_{+i}=-\tfrac12\,\vartheta(x^{+})\,\D_i\Phi$, the transverse velocity acquired at the
shock. The profile is an exact vacuum pp wave in Brinkmann form, so $R_{i+j+}$ has no nonlinear
part and $\mR_{ij}=-\,\delta(x^{+})A_{ij}$ holds exactly with $\mR^{(2)}=0$. The matrix $A_{ij}$ is
symmetric, is independent of $x^{+}$, and by \eq{SecF9} its trace is the source,
$\tr A=\tfrac12\D_{\bperp}^{2}\Phi$ and $\delta(x^{+})\tr A=R_{++}=\tfrac{\kappa}{2}\,T_{++}$, so
$\tr A=0$ away from the matter.
Since $A$ commutes with itself, \eq{SecC4} integrates at once and \eq{SecC3} is a genuine
exponential,
\beq\label{SecI9}
W_{ij}(x^{+})\,=\,\Big[\exp\big(x^{+}A\big)\Big]_{ij}\,,\qquad x^{+}>0\,,\qquad
W_{ij}\,=\,\delta_{ij}\,,\quad x^{+}<0\,.
\eeq
The geometric generator is different. With $\mR=\delta(x^{+})A$ the Riccati equation
\eq{SecG9} gives a finite jump of $\hat{s}$ at the shock and $\D_{+}\hat{s}=-\hat{s}^{2}$
after it,
\beq\label{SecI10}
\hat{s}(x^{+})\,=\,-\,A\,\big(\bone\,-\,x^{+}A\big)^{-1}\,,\qquad
\bW^{\rm J}\,=\,P\exp\!\int_{0}^{x^{+}}\!\!\hat{s}\,=\,
\bone\,-\,x^{+}A\,=\,\delta_{ij}\,-\,\frac{x^{+}}{2}\,\D_i\D_j\Phi\,,
\eeq
which is the thin lens Jacobi matrix of gravitational lensing. The generator $\hat{s}$ against
$\hat{G}$ of \eq{SecI8}, and with them $\bW^{\rm J}$ against $\bW$ of \eq{SecI9}, are opposite and
inverse at order $h$ by \eq{SecF802}, the inversion fails at order $h^{2}$,
\beq\label{SecI11}
\hat{s}\,=\,-\,\hat{G}\,-\,\D_{+}^{-1}\big(\hat{G}\hat{G}\big)\,+\,\dots\,=\,
-\,\vartheta(x^{+})\Le A\,+\,x^{+}A^{2}\,+\,\dots\Ra\,,\qquad
\bW^{\rm J}\,\bW\,=\,\bone\,-\,\tfrac12\,\big(x^{+}A\big)^{2}\,+\,O(h^{3})\,,
\eeq
which is \eq{SecG14} made explicit. The origin of the difference is the support in $x^{+}$.
The curvature $\mR=-\,\delta(x^{+})A$ acts at one instant, while its primitive
$\hat{G}=\vartheta(x^{+})A$ does not vanish afterwards, so a first order transport generated by
$\hat{G}$ keeps accumulating in the region where the metric is already flat. The Riccati
connection \eq{SecI10} decays as $1/x^{+}$ and stops. Equivalently, by \eq{SecC5} the
perturbative matrix obeys $\D_{+}^{2}W=(G+\hat{G}\hat{G})W$ with
\beq\label{SecI12}
G_{ij}\,+\,\hat{G}_{ik}\hat{G}_{kj}\,=\,\delta(x^{+})\,A_{ij}\,+\,
\vartheta(x^{+})\,\big(A^{2}\big)_{ij}\,,
\eeq
the second term being a tidal matrix in flat space. For $x^{+}>0$ geodesic deviation reads
$\D_{+}^{2}W=0$ and every solution is linear in $x^{+}$, which $\bone-x^{+}A$ is and
$\exp(x^{+}A)$ is not. The substitution $\hat{G}\to\hat{s}$ of \eq{SecF801} is exactly the
demand that the effective tidal matrix $-\mR+\hat{S}\hat{S}$ of \eq{SecG401} be replaced by the true
curvature $\mR$ of \eq{SecG9}. Note that the difference has nothing to
do with the transverse position at which the tidal matrix is taken: here $\mR$ is supported at
$x^{+}=0$, where the ray is still at its impact parameter, so that ambiguity is absent and the
whole of \eq{SecI11} is the $\hat{s}\hat{s}$ term.

The optical scalars follow from \eq{SecG22}. The eigenframe of $A$ does not depend on
$x^{+}$, so $\beta$ is constant, $\alpha=\beta$, $\Psi=0$ exactly and both matrices are symmetric: a single
shock produces no image rotation. In vacuum $\tr A=0$ and the eigenvalues of $A_{ij}$ are
$\pm a$ with $a=a(x_{\bperp})$, so that
\beq\label{SecI13}
\bW^{\rm J}:\quad
\Delta^{-1/2}=\sqrt{1-(ax^{+})^{2}}\,,\quad
\Sigma={\rm artanh}\,(ax^{+})\,,\quad
\theta=-\,\frac{2a^{2}x^{+}}{1-(ax^{+})^{2}}\,,\quad
s=\frac{a}{1-(ax^{+})^{2}}\,,
\eeq
\beq\label{SecI14}
\bW:\quad
\Delta^{1/2}=1\,,\qquad \Sigma=a\,x^{+}\,,\qquad \theta=0\,,\qquad s=a\,,
\eeq
the scalars of \eq{SecI13} being those of $\hat{s}$ and the scalars of \eq{SecI14} those of
$\hat{S}$, which for this profile is $\hat{S}=\hat{G}=\vartheta(x^{+})A$ with $\tr A=0$.
Both satisfy \eq{SecG23}: $\tr\bW^{\rm J}=2$ and $\tr\bW=2\cosh(ax^{+})$. The geometric matrix
focuses, with the expansion negative and of second order in $h$, as required by the
Raychaudhuri equation \eq{SecG11} with $\sigma$ as the only source,
$\D_{+}\theta=-\tfrac12\theta^{2}-2s^{2}$. The perturbative matrix is area preserving, has
$\theta=0$ and never focuses. For the Aichelburg and Sexl profile $\Phi=4GE\ln x_{\bperp}^{2}$
of a source of energy $E$, we correspondingly could write:
\beq\label{SecI15}
A_{ij}\,=\,\frac{4GE}{x_{\bperp}^{2}}\Le \delta_{ij}\,-\,
\frac{2\,x_{i}x_{j}}{x_{\bperp}^{2}}\Ra\,,\qquad
a\,=\,\frac{4GE}{x_{\bperp}^{2}}\,,\qquad
x^{+}_{\rm caustic}\,=\,\frac{1}{a}\,=\,\frac{x_{\bperp}^{2}}{2R_{s}}\,,
\eeq
with $R_{s}=2GE$, the eigendirections being radial and tangential and the deflection at the
shock being $4GE/|x_{\bperp}|$. By \eq{SecI13} one has $\det\bW^{\rm J}=1-(ax^{+})^{2}$, so the
caustic condition \eq{SecG2802} reads $a\,x^{+}=1$ and is the Einstein ring of the line,
\beq\label{SecI1501}
b_{E}\,=\,\sqrt{4GE\,x^{+}}\,=\,\sqrt{2R_{s}x^{+}}\,,\qquad
\Sigma\,\simeq\,\frac{4GE\,x^{+}}{b^{2}}\,,\quad \Delta\,\simeq\,1\,,\qquad b\,\gg\,b_{E}\,,
\eeq
where the ray density \eq{SecG27} diverges and the eikonal saddle point fails. At
the ring $\Lambda_{2}=1-ax^{+}\to0$ while $\Lambda_{1}=1+ax^{+}\to2$, which is the degeneration
described after \eq{SecG2801}. It belongs to $\bW^{\rm J}$ only: by \eq{SecI14} the perturbative
matrix has $\det\bW=1$ and no caustic at any $x^{+}$.

The currents differ accordingly; from \eq{SecC6} we have:
\beq\label{SecI16}
\mO^{\rm J}_{++}=2\,\D_{+}^{2}\,\tr\bW^{\rm J}\,=\,-\,2\,\delta(x^{+})\,\tr A=
-\,2\,R_{++}\,,\,\,
\mO_{++}=2\,\D_{+}^{2}\,\tr\bW=2\,R_{++}\,+\,4\,a^{2}\cosh (ax^{+})\,\vartheta(x^{+})\,,
\eeq
the second expression being written for $\tr A=0$ in the last term. The geometric current
terminates at the Born term and carries it with the opposite sign, $\bW^{\rm J}$ being the inverse
transport, while the perturbative one has the correct Born term and a tower whose first member,
$2\,\vartheta(x^{+})\tr A^{2}$, is the $h^{2}$ vertex of \eq{SecD6}. The perturbative generator
is therefore the one that carries the effective current of Section~\ref{SecD}, and the
identification of $W_{ij}$ with the Jacobi propagator is exact at order $h$ and approximate
beyond it.

For a sequence of shocks with profiles $\Phi_{n}$ at positions $x^{+}_{n}$ separated by light
cone distances $\ell_{n}$, the generator is piecewise constant and the ordering becomes
nontrivial,
\beq\label{SecI17}
\bW_{\rm out}\,=\,\prod_{n}^{\longleftarrow}\exp\big(\ell_{n}B_{n}\big)\,,\qquad
B_{n}\,=\,\sum_{m\,\leq\,n}A_{m}\,,\qquad A_{n\,ij}\,=\,\tfrac12\,\D_i\D_j\Phi_{n}\,,
\eeq
the factors being ordered with increasing $n$ to the left, as \eq{SecG4} requires, so that
$\bW_{\rm out}=e^{\ell_{N}B_{N}}\cdots e^{\ell_{1}B_{1}}$. Here $[A_{n},A_{m}]\neq0$ produces
the image rotation $\Psi$ of \eq{SecG21} at order $h^{2}$.
The geometric matrix in the same background is the product of thin lens and free drift factors,
which is the first order form of \eq{SecF801} written on the pair
$(\bW^{\rm J},\D_{+}\bW^{\rm J})$,
\beq\label{SecI18}
\begin{pmatrix} \bW^{\rm J} \\ \D_{+}\bW^{\rm J}\end{pmatrix}\Bigg|_{\rm out}\,=\,
\prod_{n}^{\longleftarrow}\begin{pmatrix} \bone & \ell_{n}\bone \\ 0 & \bone\end{pmatrix}
\begin{pmatrix} \bone & 0 \\ -\,A_{n} & \bone\end{pmatrix}
\begin{pmatrix} \bone \\ 0\end{pmatrix}\,.
\eeq
Each factor belongs to $Sp(4,\mathbb{R})$ and the product is exact. The two resummations,
\eq{SecI17} and \eq{SecI18}, are inverse at order $h$ and the inversion fails at order $h^{2}$ in the
way \eq{SecI11} describes.

%%%%%%%%%%%%%%%%%%%%%%%%%%%%%%%%%%%%%%%%%%%%%%%%%%%%%%%%%%%%%%%%%%%%%%%%%%%%%%%%

\section{Results and conclusion}\label{SecJ}

 The approaches to the descriptions of gravitational high energy scattering are developing extensively and intensively during the last decades,
see \cite{Veneziano,Veneziano1,Veneziano2,Veneziano3,Veneziano4,Veneziano5,Veneziano6,Veneziano7,Veneziano8,Veneziano9,Veneziano10} and references therein
as a small selection of published papers concerning the subject. Then, the possible Reggeization of the scattering amplitudes and application of the corresponding counterpart schemes from QCD in gravity related calculations has not been overlooked as well, see 
\cite{SabVera,SabVera1,SabVera2,SabVera3,SabVera4,Venug,Venug1,Venug2,Venug3,Venug4,Venug5,Venug6}. In this respect, the proposed article is focused on the development
of the notion of effective action introduced by L.~N.~Lipatov for the description of Reggeization in connection to the gravity high energy scattering. 

 The approach proposed in \cite{LipatovGrav1} was based on the formalism where the construction of the effective vertices of interaction of gravitational Reggeons with
the regular ones was performed perturbatively, from the bottom up. The way then is opposite to what the QCD effective action proposes. There the interactions were written ab initio in a full and closed way through the use of the P-ordered exponential, the requested vertices then could be calculated on the basis of the given non-local expression at any requested order. Therefore, besides that such a perturbative bottom-up approach is extremely involved, it is based on some assumption about the form of the effective current which a priori is not known and must be somehow guessed. An additional problem of this scheme is that in the answers a product of non-local operators appears immediately and then the final answer depends on the way of treating the uncertainties arising in such products. 

 Therefore, in this paper we proposed a way of construction of a gravitational operator, similar to some extent to the QCD Wilson line, where the P-ordering along 
the path appears automatically by the construction and not through deus ex machina. The solution is an operator $W_{ij}$ defined through \eq{SecC2}-\eq{SecC3},
it is a matrix with the transverse indices which lives on the 2D screen perpendicular, at given $x^{+}$, to the paths of the scattering objects. The requested ordering then is a consequence of the non-commutativity of the operator along the corresponding path and this is an analog of the non-Abelian gluon fields representation we have in QCD. The form of the corresponding generator depends on the used approximation, see \eq{SecC9}-\eq{SecC15}. The simplest one is that which corresponds to the high energy limit and the most complete, in light cone definition and coordinates, is that which corresponds to the linearized Riemann tensor. Then, the construction by definition is linear diffeomorphism invariant and includes all sub-leading contributions of the generally possible non-leading Reggeon fields. Moreover, the operator allows a full covariant description discussed in Section~\ref{SecG}. It is important that in the covariant form the generator in this set-up is a full Riemann tensor contracted with the vectors along the light cone paths of the scattering particles, i.e. it is a fully diffeomorphism invariant object. It is interesting to note, that this $W_{AB}$ operator can be then represented in terms of optical invariants through the irreducible expansion of the generator, see 
\eq{SecG7}. In this formulation the scattering is described through the terms of the expansion, three for each particle, the generalization of this description for the case of quantum corrections is not direct then, we postpone this task for other publications. It is worth noticing, finally, that a similar construction, after the proper redefinition of the generator of course, could be introduced in QCD, there it will reproduce a sub-eikonal correction to the usually ordered non-Abelian Wilson line.

 The main results we acquire by this construction are the effective interaction vertices between the Reggeized and regular gravitons or between two different light cone gravitons appearing in the variant of interacting truncated propagators \eq{SecE8}. It is important to underline, that the Reggeization appears as one-loop contribution to the t-channel propagator, from this point of view it does not matter which form of interaction term, \eq{SecE2},\eq{SecE3} or \eq{SecE8}, to use.
What is important is the form of the effective currents at $h^{2}$ precision, i.e. the form of the vertices appearing because of the P-ordering of the exponential. Having in mind the perturbative calculations, \cite{LipatovGrav1,MyGrav}, where the effective
current was calculated till the $h^{2}$ precision order, it turns out that in the proposed set-up there is an additional part to the current of the same order, 
\eq{SecD9}, which was not accounted for by the perturbative approach. The reason for the mismatch is that the current we work with in \cite{LipatovGrav1,MyGrav}
has a particular restricted form, see \eq{SecB2}, and these additional terms could not be fixed by this kind of calculations.
Nevertheless, the question whether these new terms modify the Reggeized graviton trajectory of \cite{LipatovGrav,MyGrav} remains open, we need a direct calculation check which will be performed in a separate publication.

 As underlined already, the trajectory of any Reggeized t-channel amplitude is defined by the form of expansion of the P-ordered exponential. Then the $h^3$ LO
order effective vertex, see \eq{SecD10}-\eq{SecD11}, determines, together with the $h^{2}$ vertex, the trajectory of the bound state of two Reggeized gravitons, i.e. the trajectory 
of the gravitational Pomeron. Continuing with this perturbative counting, each next order effective vertex determines a trajectory of the next order bound states of the Reggeized gravitons. Namely, we remind that differently from QCD, in Einstein-Hilbert gravity construction each bound state of the Reggeized gravitons,
including the single one, has the quantum numbers of the vacuum and is the same by this criterion. What separates the state then is a parameter counting, each state has its own proportionality to the Newton constant, through the coupling inside the impact factors, and its own energy dependence enhanced for the states with larger number of bounded Reggeized gravitons. Therefore, in this respect, at the $s\,\rightarrow\,\infty$ Regge limit, the maximal contribution to the non-eikonalized t-channel amplitude comes from the bound state with infinite number of Reggeized t-channel gravitons. Of course, for the finite value of the c.m. energy, the single Reggeized graviton provides a maximal contribution, which is, nevertheless, enhanced in comparison to the single regular graviton exchange, see discussion in \cite{MyGrav}. 

 Then, direct use of the proposed formalism is a calculation of the amplitudes of bound states of Reggeized gravitons, similar to what was done in \cite{MyGrav}.
The application of the gravitational scattering processes amplitudes can be found in 
\cite{Veneziano,Veneziano1,Veneziano2,Veneziano3,Veneziano4,Veneziano5,Veneziano6,Veneziano7,Veneziano8,Veneziano9,Veneziano10}. It is important that the formalism developed in 
\cite{MyGrav} allowed a description of the particles production through the gravitational interaction, it could be important for the possible phenomenological description of the data. It was clarified in \cite{OurZubkov1,Our11}, the interaction term in the overall effective action can be written and used in the three 
equivalent forms. There \eq{SecE8} carries it in the operators $\mO_{\pm\pm}$ form, \eq{SecF14} in terms of the Riemann tensor and \eq{SecH3} covariantly on the
null congruence, with $\Box_{S}$ as the Laplacian on the screen. The RFT forms \eq{SecE3} and \eq{SecF16}, with the generator built
from the Reggeon field itself, follow from the same expressions. These different formulations of the same object, as already was mentioned above, provide the same 
quantum picture at LO, beyond it the results can be different if we talk about RFT formulation, see \cite{Our13}.

 Yet, in each of these forms the interaction is purely transverse. By \eq{SecD12} the operator is a total derivative in its own light cone coordinate, the two operators of the interaction term depend on one such coordinate each separately and on the transverse ones both, and $\D_{\bperp}^{-2}$ commutes with both longitudinal integrations. Then there the $d^{4}x$ integral collapses onto a $d^{2}x_{\bperp}$ integral of boundary values, \eq{SecE9} and \eq{SecF17}.
The RFT expression \eq{SecF20} stays local in the transverse plane, the Reggeon fields not being integrated out there and we obtain a variant of 2D effective RFT.
It is interesting that the same interaction term allows a form without matrix and ordering. Written as \eq{SecG23} presents, the expression
\eq{SecH407} is a bilinear form in three scalar functions on the screen for each line, the density $\Delta$ of the ray bundle, the accumulated shear $\Sigma$ and the image rotation $\Psi$, each entering through two derivatives along the ray. The orientation $\alpha$ of the
shear drops out by the second relation of \eq{SecG2201}, so the effective current reads three of
the four parameters of the transport.

 These optical invariants deserve a separate discussion then, because
they are not an auxiliary parametrization but the standard variables of light ray optics in a
curved background, and the whole formalism can be read in that language. The decomposition
\eq{SecG7} of the generator is the irreducible splitting
of a screen tensor into the expansion, the shear and the twist. These are the optical scalars of
Sachs \cite{Sachs}, defined there for the deformation matrix of a null congruence, and the trace
equation is the Raychaudhuri equation \cite{Ray}, which appears here as \eq{SecG904} rather than
as an independent input. The shear equation \eq{SecG1101} and the twist equation \eq{SecG1102}
are the remaining Sachs equations. Two features are specific to the set-up of the formalism. First, the twist
vanishes identically for $\hat{S}_{AB}$, \eq{SecG701}, because $\hat{S}$ is the light cone
primitive of $\mR_{AB}$ and the latter is symmetric by the pair symmetry of the Riemann tensor,
so the congruence is hypersurface orthogonal. Second, the
rotation $\Psi$ of \eq{SecG21} is not a twist. It is generated by the ordering, through the
commutator \eq{SecG2101} of shears of different orientation, and it is the one parameter of the
transport with no counterpart in the Abelian eikonal phase.

 As already mentioned, the four parameters of \eq{SecG22} are in one to one correspondence with the quantities used in
gravitational lensing, where $\bW^{\rm J}$ is the Jacobi matrix of the lens mapping and $\bW$ is
its inverse, the amplification matrix, see \cite{SEF} for the systematic treatment and
\cite{SSE} for the derivation of the lens approximation from the exact light propagation
equations. In that dictionary $\Delta=\det\bW$ is the ray density on the screen, \eq{SecG27},
$\Sigma$ fixes the axis ratio $e^{2\Sigma}$ of the image, $\alpha$ its orientation and $\Psi$ its
rotation, the principal magnifications being the singular values
$\Lambda_{1,2}=\Delta^{-1/2}e^{\pm\Sigma}$ of \eq{SecG28}. The rotation of the image by a
sequence of lens planes and its second order onset are treated in \cite{PenMao}; the
spacetime formulation of the lens mapping, of the critical curves and of the Morse index is
reviewed in \cite{Perlick}.

 The scale parameter has also an independent meaning. The quantity $\Delta$ is the van Vleck and
Morette determinant of the geodesic flow \cite{VanVleck}, the object which supplies the one loop
weight of the semiclassical propagator \cite{Morette}, and $\Delta^{1/2}$ is therefore the prefactor of the
eikonal amplitude \eq{SecG29}. Its behavior at the focal points, the sign of $\det\bW^{\rm J}$
and the phase collected through a caustic are discussed in \cite{Visser} and, in the language of
bitensors and of the Hadamard construction, in \cite{PPV}. The covariant set-up of
Section~\ref{SecG}, a single null geodesic with a Sachs dyad parallel transported along it and
the curvature contracted with the tangent, is the data of the Penrose limit of the background
\cite{Penrose}, the curvature along the geodesic being what that limit retains, see \cite{Blau},
and the shock wave example of Section~\ref{SecI} is treated from that point
of view in \cite{Shore}. The translation is therefore direct in both directions, results on
focusing, on caustics and on the van Vleck determinant obtained there could be applied to the
effective current of \eq{SecG23}, and the ordering statement \eq{SecG2101}, which is proper to
the high energy problem, adds to it the parameter $\Psi$ which the single lens analysis does not
see.

 Consequently, two ordered objects appear in the formalism and we have to distinguish them. The operator
$\bW$ of \eq{SecC3} is generated by $\hat{S}_{AB}\,=\,-\D_{+}^{-1}\mR_{AB}$ and carries the
effective current. The Jacobi propagator $\bW^{\rm J}$ of \eq{SecF801} is generated by the
deformation matrix $\hat{s}_{AB}$ which solves the Riccati equation \eq{SecG9} with the curvature itself. 
They are related to two different geometrical objects, the first transports the deviation form and the second the deviation vector,
which is the position of the single transverse index; \eq{SecF802} relates them as
$\bW=\big[(\bW^{\rm J})^{T}\big]^{-1}+O(h^{2})$. The inversion fails at the second order, for
the shock profile the answer is \eq{SecI11}. We obtained therefore that the action's effective current is
built on the first object while the optical scalars, the caustics and the van Vleck determinant belong to the second.
Nevertheless, these two objects depend on the ordering notion introduced and 
the ordering has an explicit criterion in the construction. The commutator of the generators at two points of the
line is \eq{SecG2101} expression, so the expansion \eq{SecC7} is an ordered one only when the
principal axes of the tidal matrix rotate along the line. For a factorized profile  of \eq{SecG2102} the direction $\beta$ is constant and the
ordering is disappearing. These are cases of a single shock wave, its smeared version and the leading eikonal 
of two point particles profile. The minimal ordered configuration is two exchanges separated in $x^{+}$
and in $x_{\bperp}$, \eq{SecI17}, so the image rotation starts at $\Psi=O(h^{2})$ in the
transport. The counting here is in the powers of the external field, i.e. in the number of the
exchanges with the source.
In the effective current the rotation appears two orders later correspondingly, there the action has the \eq{SecG23} trace
where $\cos\Psi$ is even, so with $\Psi=O(h^{2})$ the first contribution of the rotation to $\mO_{++}$ is of the order $h^{4}$.
The same takes place in the language of the expansion: the ordered double integral differs from the
unordered one by $\tfrac12\int\!\!\int\big[\hat{G}_{1},\hat{G}_{2}\big]=\Psi\,\epsilon$ of
\eq{SecG2104}, the matrix $\epsilon$ is antisymmetric and $\tr\Le \Psi\epsilon\Ra=0$, so the
$h^{2}$ current does not see the ordering at all. With $L^{(n+1)}$ of \eq{SecE4}-\eq{SecE5}
carrying $\mO|_{h^{n}}$, the classical appearance of the ordering enters therefore by the five
Reggeon vertex and not by the single trajectory. It enters there through two structures: with
$\bW=\exp\Omega$ and the Magnus terms $\Omega_{1}$ symmetric, $\Omega_{2}=\Psi\epsilon$
antisymmetric and $\Omega_{3}$ symmetric traceless, one has
$\tr\bW|_{h^{4}}\supset\tfrac12\tr\Omega_{2}^{2}+\tr\Le \Omega_{1}\Omega_{3}\Ra=-\Psi^{2}+
\tr\Le \Omega_{1}\Omega_{3}\Ra$, while $\tr\Omega_{4}$ and $\tr\Le \Omega_{1}^{2}\Omega_{2}\Ra$
vanish, so the second structure is the ordering carried by the accumulated shear. The counting does not apply to the induced current
\eq{AppB1601}, where the structure $W_{li}\hat{W}_{ik}$ stands outside the trace and its
antisymmetric part survives.
The same statement is known in the lensing literature, where one deflector shears an image but
cannot rotate it, see \cite{PenMao}. Of course, here we discussed the classical results of the ordering, the Reggeization is a quantum effect 
and request the P-ordered operator from the very beginning.

 Interesting and important question we discussed is also a comparison of the present construction with regular and Abelian gravitational Wilson lines,
it is done in Section~\ref{SecH}. The two exponentials are the Abelian phase, the regular Wilson line, and the Hessian of the same eikonal, the operator of the formalism, with the deflection between them as the first moment, as
\eq{SecH11} shows in one line. They combine into a single $GL(2,\mathbb{C})$ transport
$\bU=e^{\,i\chi}W$ of \eq{SecH1001}, the phase generator being a multiple of the unit matrix and
leaving the ordering untouched. The Abelian factor carries the longitudinal information, the
Shapiro delay $\Delta x^{-}=-\chi/p^{+}$, which $W_{ij}$ does not have, and the matrix factor
carries the shear and the focusing, which the phase does not have. Generally, the
construction fixes also where the eikonal description stops. The amplitude carries the prefactor
$\Delta^{1/2}$ and the phase $e^{-i\pi n_{c}/2}$ of \eq{SecG29}, with $n_{c}$ the number of
caustics on the segment counted with multiplicity. The caustics are the zeros of
$\det\bW^{\rm J}$, \eq{SecG2802}, and they belong to the Jacobi matrix alone. The ordered
exponential cannot produce them, $\det\bW=\exp\int^{\lambda}\!d\lambda'\,\theta$ is a real
exponential with $\theta$ finite, so $\bW$ stays in $GL^{+}(2,\mathbb{R})$ and never degenerates.
For the Aichelburg and Sexl profile the condition reads $a\,x^{+}=1$ and gives the Einstein ring
$b_{E}=\sqrt{4GE\,x^{+}}$ of \eq{SecI1501}, which is the impact parameter where the ray density
diverges and the saddle point of \eq{SecH10} fails.

 For the classical equations of motion we need an EMT produced by the interaction Lagrangian. It can be found by the
variation of the operator with respect to the metric and it is calculated in
Appendix~\ref{AppB}. The answer is obtained through the pair $W$ and $\hat{W}$ of \eq{AppB1}, with
$\hat{W}(z^{+})W(z^{+})=W(+\infty)$ of \eq{AppB3} replacing the Wronskian relation of the second
order form, and the induced current of the classical equations of motion is \eq{AppB1601}. Its
first term is of the second perturbative order in $h$. The three forms
\eq{AppB11}-\eq{AppB13} of the vertex operator $\mV^{\mu\nu}_{kl}$ correspond to the three
generators \eq{SecC11}, \eq{SecC12} and \eq{SecC14}, so the current is available in the same
three approximations as the operator itself. The expression then is invariant with respect to the diffeomorphisms restricted by 
the forms of \eq{AppB12}-\eq{AppB13} generators, the last one then provides the invariance with respect to the linearized diffeomorphisms.

 The last remark we have is about the accuracy of the results and open points left. The boundary expressions
\eq{SecE10} and \eq{SecF18} require a regularization: the connection $\hat{G}_{ij}$ does not
vanish at the ends of the line. With the retarded primitive its integral grows linearly and $W(+\infty)$ diverges
already for the shock wave profile. With the symmetric $\D_{\pm}^{-1}$ adopted after \eq{SecC8} the same integral is finite,
$\int dx^{+}\hat{G}_{ij}=-\int dx^{+}x^{+}G_{ij}$, but it depends on the origin of the light cone coordinate, a shift
$x^{+}\rightarrow x^{+}\,+\,a$ changing it by $a\bar{G}_{ij}$. In any case, the total transport is fixed only together with a
prescription. The $\Sigma$ and the
$\Psi$ of \eq{SecH407} are defined by the polar decomposition \eq{SecG20} of the ordered
exponential and not by the generator at a point, so their extraction from $\hat{S}$ is the
ordering problem again, and the Magnus split of the two factors is fixed only up to $O(h^{3})$.
The correspondence with the Jacobi propagator, and with it the whole dictionary of the ray
bundle, holds at the first order in $h$ and fails at the second, \eq{SecF802} and \eq{SecI11}.
The transverse component $h_{ij}$ is dropped by \eq{SecC15} as the least contributing one in the
Regge limit. In general this ray optic construction is classical, the extension to the quantum corrections is not
immediate and it is postponed, and, as mentioned, the check of the Reggeized graviton trajectory against the
additional terms of \eq{SecD9} is postponed as well.

 In conclusion, we hope that the proposed formalism for describing high-energy gravitational scattering processes will be useful for the development of gravitational theory in both its classical and quantum aspects.

%%%%%%%%%%%%%%%%%%%%%%%%%%%%%%%%%%%%%%%%%%%%%%%%%%%%%%%%%%%%%%

\section*{Declaration of use of generative Anthropic Claude Opus AI engine}

 During the preparation of the manuscript, the author used the Anthropic Claude Opus 5.5 AI software engine to check symbolic computations, 
to check grammar and text formatting.
After using this tool, the author reviewed and edited the content as needed and bears full responsibility for the content of the published article.

%%%%%%%%%%%%%%%%%%%%%%%%%%%%%%%%%%%%%%%%%%%%%%%%%%%%%%%%%%%%%%%%%%%%%%%%%%%%%%%%

\appendix
\newpage
\section{Algebraic identities}\label{AppA}
\renewcommand{\theequation}{A.\arabic{equation}}
\setcounter{equation}{0}

For the simplification of the calculation algebra we use the quantities defined as follows:
\beq\label{AppA1}
v=\D_{+}^{-1}h_{++}\,,\,\,\, w=\D_{+}^{-2}h_{++}\,,\,\,\, K_{ij}=\D_i\D_j h_{++}\,,\,\,\, U_{ij}=\D_i\D_j w\,. 
\eeq
We have then
\beq\label{AppA2}
K_{ik}U_{ki}+\big(\D_i\D_k v\big)^{2}
=\D_{+}\Big[\big(\D_i\D_k v\big)\big(\D_i\D_k w\big)\Big].
\eeq
The proof is simple: 
\beq\label{AppA3}
\D_{+}\big[(\D_{+}U_{ik})U_{ki}\big]\,=\,(\D_{+}^{2}U_{ik})U_{ki}\,+\,(\D_{+}U_{ik})(\D_{+}U_{ki})\,;\,\,\,
\D_{+}U_{ik}\,=\,\D_i\D_k v\,,\,\,\,\D_{+}^{2}U_{ik}\,=\,K_{ik}\,.
\eeq
 The last term is $(\D_i\D_k v)^{2}$ because $v$ is a scalar, so $\D_i\D_k v$ is symmetric.
Further we have:
\beq\label{AppA4}
\tfrac14\,\D_{\bperp}^{2}\big(\D_k v\,\D_k v\big)
=\tfrac12\big(\D_i\D_k v\big)^{2}+\tfrac12\,\D_k v\,\D_k\big(\D_{\bperp}^{2}v\big).
\eeq
Here we have: 
\beq\label{AppA5}
\D_i\D_i(\D_k v\,\D_k v)=2\D_i(\D_i\D_k v\,\D_k v)\,=\,2(\D_i\D_k v)^{2}\,+\,2\D_k v\,\D_k\D_{\bperp}^{2}v\,.
\eeq
The last identity is simple, from \eq{SecF1} we obtain
\beq\label{AppA6}
2\D_{+}^{-2}R_{i+j+}\,=\,\D_{+}^{-1}\Le \D_j h_{i+}\,+\,\D_i h_{+j}\Ra\,-\,h_{ij}\,-\,\D_{+}^{-2}\D_i\D_j h_{++}
\,=\,Y_{ij}-h_{ij}\,.
\eeq

\section{Variation of the effective current}\label{AppB}
\renewcommand{\theequation}{B.\arabic{equation}}
\setcounter{equation}{0}

 The equations of motion require the variation of $\mO_{++}$ of \eq{SecC6} with respect to $h$.
By definition the answer, with the Reggeon field included, see \eq{SecE1}, is the
energy-momentum tensor of the given action. Similarly to what is done in \cite{Our,Our01}, we
introduce two $W$ operators, the regular one and the backward ordered transport along the same line;
by \eq{AppB3} further it defined as $\hat W=W(+\infty)W^{-1}$. So, consider \eq{SecC4},
which is of the first order:
\beq\label{AppB1}
\D_{+}W_{ij}=\hat G_{ik}W_{kj},\;\; W\big|_{x^{+}\to-\infty}=\bone;
\qquad
\D_{+}\hat W_{ij}=-\,\hat W_{ik}\hat G_{kj},\;\; \hat W\big|_{x^{+}\to+\infty}=\bone,
\eeq
that is
\beq\label{AppB2}
W(z^{+})=P\exp\int_{-\infty}^{z^{+}}\!\!dx^{+}\,\hat{\bG}(x^{+}),
\qquad
\hat W(z^{+})=P\exp\int_{z^{+}}^{+\infty}\!\!dx^{+}\,\hat{\bG}(x^{+}),
\eeq
the counterparts of the operators $O$ and $O^{T}$ of \cite{Our} with the gauge field replaced
by $\hat G_{ij}$, where 
\beq\label{AppB3}
\hat W(z^{+})\,W(z^{+})\,=\,W(+\infty)
\eeq
for every $z^{+}$, which replaces the Wronskian relation of the second order form.
Now we borrow the derivation from \cite{Our}, redefining the notation introduced there in order
to avoid confusion. Writing the QCD counterparts of the operators on the left hand side of the
expressions, we have, for the function in the exponent of the ordered exponential, which there
was the $+$ component of the gluon field:
\beq\label{AppB4}
v_{+}\,\rightarrow\,\hat{G}_{ij}(x)\,=\,\hat G_{+x}\,;
\eeq
for the Green's function of the $\D_{+}$ operator:
\beq\label{AppB5}
G_{xy}^{+}\,\rightarrow\,F^{+}_{xy}\,=\,F^{+0}_{xy}\,+\,F^{+0}_{xz}\,\hat G_{+z}\,F^{+}_{zy}\,,\,\,\,\,\D_{+x}F^{+0}_{xy}\,=\,\delta_{xy}\,;
\eeq
and for the operators themselves:
\beq\label{AppB6}
O_{x}\,\rightarrow\,W_{x}\,,\,\,\,O_{x}^{T}\,\rightarrow\,\hat W_{x}\,,
\eeq
as was mentioned above; see further details in \cite{Our}. Now, because we have here the same
ordered object, all the calculations of \cite{Our} can be translated directly to the case of the
gravitational operator, we obtain then:
\beq\label{AppB7}
\D_{+x} \delta W_{x}\,=\,\hat W_{x}\,\delta\hat G_{+x}\,W_{x}\,
\eeq
or
\beq\label{AppB701}
\D_{+} \delta W_{ij}(x^{+})\,=\,\hat W_{ik}(x^{+})\,\delta\hat G_{kl}(x^{+})\,W_{lj}(x^{+})\,.
\eeq
Taking the trace we obtain in turn
\beq\label{AppB8}
\D_{+} \tr\,\delta W(x^{+})\,=\,\hat W_{ik}(x^{+})\,\delta\hat G_{kl}(x^{+})\,W_{li}(x^{+})\,=\,W_{li}(x^{+})\,\hat W_{ik}(x^{+})\,\delta\hat G_{kl}(x^{+})\,
\eeq
or 
\beq\label{AppB9}
\frac{\D_{+}\delta\, \tr\, W(x^{+})\,}{\delta \hat G_{st}(x^{+})}\,=\,\frac{1}{2}\Le W_{si}(x^{+})\,\hat W_{it}(x^{+})\,+\,W_{ti}(x^{+})\,\hat W_{is}(x^{+})\Ra\,.
\eeq
For our purpose we introduce the expansion
\beq\label{AppB10}
\delta G_{kl}(x^{+})\,=\,\mV_{kl}^{\mu\nu}(x^{+})\,\delta h_{\mu\nu}(x^{+})\,,
\eeq
where $\mV_{kl,\mu\nu}$ is a differential operator and the pair $\mu\nu$ is contracted with the
metric. The three generators of \eq{SecC11}, \eq{SecC12} and \eq{SecC14} give its three forms,
\beqar\label{AppB11}
\delta G^{(K)}_{ij}\,=\,\mV^{(K)}_{ij,\mu\nu}\,\delta h_{\mu\nu}\,&=&\,\tfrac12\,\D_i\D_j\,\delta h_{++}\,;
\\
\delta G^{(Y)}_{ij}\,=\,\mV^{(Y)}_{ij,\mu\nu}\,\delta h_{\mu\nu}\,&=&\,
-\,\tfrac12\,\D_{+}\Le \D_i\,\delta h_{+j}\,+\,\D_j\,\delta h_{+i}\Ra\,+\,\tfrac12\,\D_i\D_j\,\delta h_{++}\,;
\label{AppB12}
\\
\delta G^{(R)}_{ij}\,=\,\mV^{(R)}_{ij,\mu\nu}\,\delta h_{\mu\nu}\,&=&\,-\,R_{i+j+}[\delta h]\,=\,
-\,\tfrac12\Le \D_{+}\D_j\,\delta h_{i+}\,+\,\D_{+}\D_i\,\delta h_{+j}\,-\,
\D_{+}^{2}\,\delta h_{ij}-\D_i\D_j\,\delta h_{++}\Ra\,,
\label{AppB13}
\eeqar
and we have
\beq\label{AppB14}
\delta \hat{G}_{ij}(x^{+},x_{\bperp})\,=\,\D_{+}^{-1}\,\delta\,G_{ij}\,=\,
\int^{x^{+}}\!\!dx_{1}^{+}\;\delta G_{ij}(x_{1}^{+},x_{\bperp})
\eeq
obtaining, for the variation of the action,
\beq\label{AppB15}
\D_{+} \tr\,\delta W(x^{+})\,=\,\hat W_{ik}(x^{+})\,\delta\hat G_{kl}(x^{+})\,W_{li}(x^{+})\,=\,W_{li}(x^{+})\,\hat W_{ik}(x^{+})\,
\D_{+}^{-1}\,\Le \mV_{kl}^{\mu\nu}(x^{+})\,\delta h_{\mu\nu}(x^{+})\Ra\,.
\eeq
Then the induced current of the classical equations of motion can be obtained directly from the \eq{SecC6} expression
\beq\label{AppB16}
\delta \mO_{++}\,=\,2\,\Big[\D_{+}\Le \,W_{li}(x^{+})\,\hat W_{ik}(x^{+})\Ra\,
\D_{+}^{-1}\,\Le \mV_{kl}^{\mu\nu}(x^{+})\,\delta h_{\mu\nu}(x^{+})\Ra\,+\,
W_{li}(x^{+})\,\hat W_{ik}(x^{+})\,\Le \mV_{kl}^{\mu\nu}(x^{+})\,\delta h_{\mu\nu}(x^{+})\Ra\Big]\,,
\eeq
the overall factor being the $c\,=\,2$ of \eq{SecD5}, since $\mO_{++}\,=\,c\,\D_{+}^{2}\tr\bW$ by \eq{SecC6}.
The induced current is the coefficient of $\delta h_{\mu\nu}$,
\beq\label{AppB1601}
j^{ind\,\mu\nu}_{++}\,=\,\frac{\delta \mO_{++}}{\delta h_{\mu\nu}}\,=\,2\,\Big[\D_{+}\Le \,W_{li}\,\hat W_{ik}\Ra\,
\D_{+}^{-1}\,\mV_{kl}^{\mu\nu}\,+\,W_{li}\,\hat W_{ik}\,\mV_{kl}^{\mu\nu}\Big]\,,
\eeq
understood as an operator acting to the right.
We notice that the first term of the current is of the second perturbative order with respect to $h$.

%%%%%%%%%%%%%%%%%%%%%%%%%%%%%%%%%%%%%%%%%%%%%%%%%%%%%%%%%%%%%%%%%%%%%%%%%%%%%%%%%%%%%%%%%%%%%

%%%%%%%%%%%%%%%%%%%%%%%%%%%%%%%%%%%%%%%%%%%%%%%%%%%%%%%%%%%%%%%%%%%%%%%%%%%%%%%%%%%%%%%%%%%%%


\newpage
\begin{thebibliography}{9}

\bibitem{Regge}
T.~Regge,
  %``Bound states, shadow states and Mandelstam representation,''
  Nuovo Cim.\ {\bf 18}, 947 (1960).

%%%%%%%%%%%%%%%%%%%%%%%%%%%%%%%%%%%%%%%%%%%%%%%%%%%%%%%%%%%%%%%%%%%%%%%%%%%%%%%%%%%%%%
\bibitem{BFKL}
L.~N.~Lipatov,
  %``Reggeization Of The Vector Meson And The Vacuum Singularity In Nonabelian Gauge Theories,''
  Sov.\ J.\ Nucl.\ Phys.\ {\bf 23}, 338 (1976).

\bibitem{BFKL1}
E.~A.~Kuraev, L.~N.~Lipatov and V.~S.~Fadin,
  %``The Pomeranchuk Singularity In Nonabelian Gauge Theories,''
  Sov.\ Phys.\ JETP {\bf 45}, 199 (1977).

\bibitem{BFKL2}
I.~I.~Balitsky and L.~N.~Lipatov,
  %``The Pomeranchuk Singularity In Quantum Chromodynamics,''
  Sov.\ J.\ Nucl.\ Phys.\ {\bf 28}, 822 (1978).

%%%%%%%%%%%%%%%%%%%%%%%%%%%%%%%%%%%%%%%%%%%%%%%%%%%%%%%%%%%%%%%%%%%%%%%%%%%%%%%%%%%%%%
\bibitem{LipatovEff}
L.~N.~Lipatov,
  %``Gauge invariant effective action for high-energy processes in QCD,''
  Nucl.\ Phys.\ B {\bf 452}, 369 (1995),
  [arXiv:hep-ph/9502308 [hep-ph]].

\bibitem{LipatovEff1}
L.~N.~Lipatov,
  %``Small x physics in perturbative QCD,''
  Phys.\ Rept.\ {\bf 286}, 131 (1997),
  [arXiv:hep-ph/9610276 [hep-ph]].

\bibitem{LipatovEff2}
L.~N.~Lipatov,
  %``Small x physics in perturbative QCD,''
  Phys.\ Rept.\ {\bf 286}, 131 (1997),
  [arXiv:hep-ph/9610276 [hep-ph]].

%%%%%%%%%%%%%%%%%%%%%%%%%%%%%%%%%%%%%%%%%%%%%%%%%%%%%%%%%%%%%%%%%%%%%%%%%%%%%%%%%%%%%%
\bibitem{Our}
S.~Bondarenko, L.~Lipatov and A.~Prygarin,
  %``Effective action for reggeized gluons, classical gluon field of relativistic color charge and color glass condensate approach,''
  Eur.\ Phys.\ J.\ C {\bf 77}, 527 (2017),
  [arXiv:1706.00278 [hep-ph]].

\bibitem{Our01}
S.~Bondarenko, L.~Lipatov, S.~Pozdnyakov and A.~Prygarin,
  %``One loop light-cone QCD, effective action for reggeized gluons and QCD RFT calculus,''
  Eur.\ Phys.\ J.\ C {\bf 77}, 630 (2017).

%%%%%%%%%%%%%%%%%%%%%%%%%%%%%%%%%%%%%%%%%%%%%%%%%%%%%%%%%%%%%%%%%%%%%%%%%%%%%%%%%%%%%%
\bibitem{Gribov}
V.~N.~Gribov,
  %``A Reggeon Diagram Technique,''
  Sov.\ Phys.\ JETP {\bf 26}, 414 (1968).
%%%%%%%%%%%%%%%%%%%%%%%%%%%%%%%%%%%%%%%%%%%%%%%%%%%%%%%%%%%%%%%%%%%%%%%%%%%%%%%%%%%%%%


\bibitem{LipatovGrav}
L.~N.~Lipatov,
  %``GRAVITON REGGEIZATION,''
  Phys.\ Lett.\ B {\bf 116}, 411 (1982).

\bibitem{LipatovGrav01}
L.~N.~Lipatov,
  %``Multi - Regge Processes in Gravitation,''
  Sov.\ Phys.\ JETP {\bf 55}, 582 (1982).

\bibitem{LipatovGrav02}
L.~N.~Lipatov,
  %``High-energy scattering in QCD and in quantum gravity and two-dimensional field theories,''
  Nucl.\ Phys.\ B {\bf 365}, 614 (1991).

\bibitem{LipatovGrav03}
L.~N.~Lipatov,
  %``Effective actions for high energy processes in QCD and in quantum gravity,''
  Subnucl.\ Ser.\ {\bf 49}, 131 (2013).

\bibitem{LipatovGrav04}
L.~N.~Lipatov,
  %``Euler-Lagrange equations for high energy effective actions in QCD and in gravity,''
  Int.\ J.\ Mod.\ Phys.\ Conf.\ Ser.\ {\bf 39}, 1560082 (2015).

\bibitem{LipatovGrav05}
L.~N.~Lipatov,
  %``Euler-Lagrange equations for the Gribov reggeon calculus in QCD and in gravity,''
  Int.\ J.\ Mod.\ Phys.\ A {\bf 31}, 1645011 (2016).

\bibitem{LipatovGrav06}
L.~N.~Lipatov,
  %``Euler-Lagrange equations for high energy actions in QCD and in gravity,''
  EPJ Web Conf.\ {\bf 125}, 01010 (2016).
%%%%%%%%%%%%%%%%%%%%%%%%%%%%%%%%%%%%%%%%%%%%%%%%%%%%%%%%%%%%%%%%%%%%%%%%%%%%%%%%%%%%%%


\bibitem{LipatovGrav1}
L.~N.~Lipatov,
  %``Effective action for the Regge processes in gravity,''
  Phys.\ Part.\ Nucl.\ {\bf 44}, 391 (2013),
  [arXiv:1105.3127 [hep-th]].

%%%%%%%%%%%%%%%%%%%%%%%%%%%%%%%%%%%%%%%%%%%%%%%%%%%%%%%%%%%%%%%%%%%%%%%%%%%%%%%%%%%%%%
\bibitem{LipatovDL}
J.~Bartels, L.~N.~Lipatov and A.~Sabio Vera,
  %``Double-logarithms in Einstein-Hilbert gravity and supergravity,''
  JHEP {\bf 07}, 056 (2014),
  [arXiv:1208.3423 [hep-th]].

%%%%%%%%%%%%%%%%%%%%%%%%%%%%%%%%%%%%%%%%%%%%%%%%%%%%%%%%%%%%%%%%%%%%%%%%%%%%%%%%%%%%%%
\bibitem{Fadin}
V.~S.~Fadin, R.~Fiore and A.~Papa,
  %``On the coordinate representation of NLO BFKL,''
  Nucl.\ Phys.\ B {\bf 769}, 108 (2007),
  [arXiv:hep-ph/0612284 [hep-ph]].

\bibitem{Fadin1}
V.~S.~Fadin and R.~Fiore,
  %``Quark contribution to the gluon-gluon - reggeon vertex in QCD,''
  Phys.\ Lett.\ B {\bf 294}, 286 (1992).

\bibitem{Fadin2}
V.~S.~Fadin, R.~Fiore and A.~Quartarolo,
  %``Radiative corrections to quark quark reggeon vertex in QCD,''
  Phys.\ Rev.\ D {\bf 50}, 2265 (1994),
  [arXiv:hep-ph/9310252 [hep-ph]].

\bibitem{Fadin3}
V.~S.~Fadin, R.~Fiore and A.~Papa,
  %``One loop Reggeon-Reggeon gluon vertex at arbitrary space-time dimension,''
  Phys.\ Rev.\ D {\bf 63}, 034001 (2001),
  [arXiv:hep-ph/0008006 [hep-ph]].

\bibitem{Fadin4}
V.~S.~Fadin and R.~Fiore,
  %``Calculation of Reggeon vertices in QCD,''
  Phys.\ Rev.\ D {\bf 64}, 114012 (2001),
  [arXiv:hep-ph/0107010 [hep-ph]].

\bibitem{Fadin5}
V.~S.~Fadin, M.~G.~Kozlov and A.~V.~Reznichenko,
  %``Radiative corrections to QCD amplitudes in quasimulti - Regge kinematics,''
  Phys.\ Atom.\ Nucl.\ {\bf 67}, 359 (2004),
  [arXiv:hep-ph/0302224 [hep-ph]].

\bibitem{Fadin6}
M.~G.~Kozlov, A.~V.~Reznichenko and V.~S.~Fadin,
  %``Impact factor for gluon production in multi-Regge kinematics in the next-to-leading order,''
  Phys.\ Atom.\ Nucl.\ {\bf 75}, 850 (2012).

%%%%%%%%%%%%%%%%%%%%%%%%%%%%%%%%%%%%%%%%%%%%%%%%%%%%%%%%%%%%%%%%%%%%%%%%%%%%%%%%%%%%%%
\bibitem{Our1}
S.~Bondarenko and S.~S.~Pozdnyakov,
  %``NNLO classical solution for Lipatov's effective action for reggeized gluons,''
  Phys.\ Lett.\ B {\bf 783}, 207 (2018).

\bibitem{Our11}
S.~Bondarenko and S.~Pozdnyakov,
  %``On correlators of Reggeon fields and operators of Wilson lines in high energy QCD,''
  Int.\ J.\ Mod.\ Phys.\ A {\bf 33}, 1850204 (2018).

\bibitem{Our12}
S.~Bondarenko and S.~Pozdnyakov,
  %``On QCD RFT corrections to the propagator of reggeized gluons,''
  Nucl.\ Phys.\ B {\bf 951}, 114854 (2020).

%%%%%%%%%%%%%%%%%%%%%%%%%%%%%%%%%%%%%%%%%%%%%%%%%%%%%%%%%%%%%%%%%%%%%%%%%%%%%%%%%%%%%%
\bibitem{OurZubkov1}
S.~Bondarenko and M.~A.~Zubkov,
  %``The dimensionally reduced description of the high energy scattering and the effective action for the reggeized gluons,''
  Eur.\ Phys.\ J.\ C {\bf 78}, 617 (2018).

\bibitem{OurZubkov2}
S.~Bondarenko, S.~Pozdnyakov and M.~A.~Zubkov,
  %``High energy scattering in Einstein\textendash{}Cartan gravity,''
  Eur.\ Phys.\ J.\ C {\bf 81}, 613 (2021).

%%%%%%%%%%%%%%%%%%%%%%%%%%%%%%%%%%%%%%%%%%%%%%%%%%%%%%%%%%%%%%%%%%%%%%%%%%%%%%%%%%%%%%%%%%

\bibitem{Ivanov}
A.~V.~Ivanov and N.~V.~Kharuk,
  %``Ordered exponential and its features in Yang{\textendash}Mills effective action,''
  Commun.\ Theor.\ Phys.\ {\bf 75}, no.8, 085202 (2023).

%%%%%%%%%%%%%%%%%%%%%%%%%%%%%%%%%%%%%%%%%%%%%%%%%%%%%%%%%%%%%%%%%%%%%%%%%%%%%%%%%%%%%%%%%%%%%%%%%%%

\bibitem{Alessio}
F.~Alessio, V.~Del Duca, R.~Gonzo and E.~Rosi,
  %``Gravitational amplitudes in the Regge limit: waveforms, shock waves and unitarity cuts,''
  [arXiv:2601.21687 [hep-th]].


%%%%%%%%%%%%%%%%%%%%%%%%%%%%%%%%%%%%%%%%%%%%%%%%%%%%%%%%%%%%%%%%%%%%%%%%%%%%%%%%%%%%%%%%%%%%%%%%%%%%%%%%%%%%%%%%%
\bibitem{MyGrav}
S.~Bondarenko,
  %``Graviton reggeization and high energy gravitational scattering of~scalar particles,''
  Eur.\ Phys.\ J.\ C {\bf 84}, 1031 (2024).

%%%%%%%%%%%%%%%%%%%%%%%%%%%%%%%%%%%%%%%%%%%%%%%%%%%%%%%%%%%%%%%%%%%%%%%%%%%%%%%%%%%%%%%%%%%%%%%%%%%%%%%%%%%%%%%%%
\bibitem{BalOper}
I.~Balitsky,
  %``Operator expansion for high-energy scattering,''
  Nucl.\ Phys.\ B {\bf 463}, 99 (1996),
  [arXiv:hep-ph/9509348 [hep-ph]].

%%%%%%%%%%%%%%%%%%%%%%%%%%%%%%%%%%%%%%%%%%%%%%%%%%%%%%%%%%%%%%%%%%%%%%%%%%%%%%%%%%%%%%%%%%%%%%%%%%%%%%%%%%%%%%%%%%

\bibitem{PenMao}
U.~L.~Pen and S.~Mao,
  %``Rotation in Gravitational Lenses,''
  Mon.\ Not.\ Roy.\ Astron.\ Soc.\ {\bf 367}, 1543 (2006),
  [arXiv:astro-ph/0506053 [astro-ph]].

%%%%%%%%%%%%%%%%%%%%%%%%%%%%%%%%%%%%%%%%%%%%%%%%%%%%%%%%%%%%%%%%%%%%%%%%%%%%%%%%%%%%%%%%%%%%%

\bibitem{NS}
S.~G.~Naculich and H.~J.~Schnitzer,
  %``Eikonal methods applied to gravitational scattering amplitudes,''
  JHEP {\bf 05}, 087 (2011),
  [arXiv:1101.1524 [hep-th]].

\bibitem{White}
C.~D.~White,
  %``Factorization Properties of Soft Graviton Amplitudes,''
  JHEP {\bf 05}, 060 (2011),
  [arXiv:1103.2981 [hep-th]].

\bibitem{White1}
D.~J.~Miller and C.~D.~White,
  %``The gravitational cusp anomalous dimension from AdS space,''
  Phys.\ Rev.\ D {\bf 85}, 104034 (2012),
  [arXiv:1201.2358 [hep-th]].

\bibitem{MNSW}
S.~Melville, S.~G.~Naculich, H.~J.~Schnitzer and C.~D.~White,
  %``Wilson line approach to gravity in the high energy limit,''
  Phys.\ Rev.\ D {\bf 89}, 025009 (2014),
  [arXiv:1306.6019 [hep-th]].
%%%%%%%%%%%%%%%%%%%%%%%%%%%%%%%%%%%%%%%%%%%%%%%%%%%%%%%%%%%%%%%%%%%%%%%%%%%%%%%%%%%%%%%%%%%%%%%%%


\bibitem{BK}
M.~Beneke and G.~Kirilin,
  %``Soft-collinear gravity,''
  JHEP {\bf 09}, 066 (2012),
  [arXiv:1207.4926 [hep-ph]].

\bibitem{BK1}
M.~Beneke, P.~Hager and R.~Szafron,
  %``Soft-collinear gravity beyond the leading power,''
  JHEP {\bf 03}, 080 (2022),
  [erratum: JHEP {\bf 04}, 141 (2024)],
  [arXiv:2112.04983 [hep-ph]].

%%%%%%%%%%%%%%%%%%%%%%%%%%%%%%%%%%%%%%%%%%%%%%%%%%%%%%%%%%%%%%%%%%%%%%%%%%%%%%%%%%%%%%%%%%%%%

\bibitem{GWL1}
D.~Bonocore, A.~Kulesza and J.~Pirsch,
  %``Classical and quantum gravitational scattering with Generalized Wilson Lines,''
  JHEP {\bf 03}, 147 (2022),
  [arXiv:2112.02009 [hep-th]].

\bibitem{GWL3}
D.~Bonocore, A.~Kulesza and J.~Pirsch,
  %``Generalized Wilson lines and the gravitational scattering of spinning bodies,''
  JHEP {\bf 05}, 034 (2025),
  [arXiv:2412.16049 [hep-th]].

\bibitem{GWL2}
K.~Fernandes, F.~L.~Lin and C.~D.~White,
  %``Double soft graviton factors from the gravitational Wilson line,''
  JHEP {\bf 02}, 135 (2026),
  [arXiv:2511.05455 [hep-th]].

\bibitem{Meggio}
P.~V.~Landshoff and O.~Nachtmann,
  %``Vacuum Structure and Diffraction Scattering,''
  Z.\ Phys.\ C {\bf 35}, 405 (1987).

\bibitem{Meggio1}
O.~Nachtmann,
  %``Considerations concerning diffraction scattering in quantum chromodynamics,''
  Annals Phys.\ {\bf 209}, 436 (1991).

\bibitem{Meggio2}
E.~Meggiolaro,
  %``A remark on the high-energy quark-quark scattering,''
  Phys.\ Rev.\ D {\bf 53}, 3835 (1996),
  [arXiv:hep-th/9506043 [hep-th]].

\bibitem{Meggio3}
O.~Nachtmann,
  %``High-energy collisions and nonperturbative QCD,''
  Lect.\ Notes Phys.\ {\bf 479}, 49 (1997),
  [arXiv:hep-ph/9609365 [hep-ph]].

\bibitem{Meggio4}
E.~Meggiolaro,
  %``Eikonal propagators and high-energy parton parton scattering in gauge theories,''
  Nucl.\ Phys.\ B {\bf 602}, 261 (2001),
  [arXiv:hep-ph/0009261 [hep-ph]].

\bibitem{Meggio5}
E.~Laenen, G.~Stavenga and C.~D.~White,
  %``Path integral approach to eikonal and next-to-eikonal exponentiation,''
  JHEP {\bf 03}, 054 (2009),
  [arXiv:0811.2067 [hep-ph]].

%%%%%%%%%%%%%%%%%%%%%%%%%%%%%%%%%%%%%%%%%%%%%%%%%%%%%%%%%%%%%%%%%%%%%%%%%%%%%%%%%%%%%%
\bibitem{Shore}
G.~M.~Shore,
  %``Memory, Penrose Limits and the Geometry of Gravitational Shockwaves and Gyratons,''
  JHEP {\bf 12}, 133 (2018),
  [arXiv:1811.08827 [hep-th]].
%%%%%%%%%%%%%%%%%%%%%%%%%%%%%%%%%%%%%%%%%%%%%%%%%%%%%%%%%%%%%%%%%%%%%%%%%%%%%%%%%%%%%%%%%

\bibitem{Veneziano}
D.~Amati, M.~Ciafaloni and G.~Veneziano,
  %``Effective action and all order gravitational eikonal at Planckian energies,''
  Nucl.\ Phys.\ B {\bf 403}, 707 (1993).

\bibitem{Veneziano1}
M.~Fabbrichesi, R.~Pettorino, G.~Veneziano and G.~A.~Vilkovisky,
  %``Planckian energy scattering and surface terms in the gravitational action,''
  Nucl.\ Phys.\ B {\bf 419}, 147 (1994),
  [arXiv:hep-th/9309037 [hep-th]].

\bibitem{Veneziano2}
D.~Amati, M.~Ciafaloni and G.~Veneziano,
  %``Towards an S-matrix description of gravitational collapse,''
  JHEP {\bf 02}, 049 (2008),
  [arXiv:0712.1209 [hep-th]].

\bibitem{Veneziano3}
M.~Ciafaloni, D.~Colferai and G.~Veneziano,
  %``Infrared features of gravitational scattering and radiation in the eikonal approach,''
  Phys.\ Rev.\ D {\bf 99}, 066008 (2019),
  [arXiv:1812.08137 [hep-th]].

\bibitem{Veneziano4}
A.~Addazi, M.~Bianchi and G.~Veneziano,
  %``Soft gravitational radiation from ultra-relativistic collisions at sub- and sub-sub-leading order,''
  JHEP {\bf 05}, 050 (2019),
  [arXiv:1901.10986 [hep-th]].

\bibitem{Veneziano5}
P.~Di Vecchia, C.~Heissenberg, R.~Russo and G.~Veneziano,
  %``The eikonal approach to gravitational scattering and radiation at $ \mathcal{O} $(G$^{3}$),''
  JHEP {\bf 07}, 169 (2021),
  [arXiv:2104.03256 [hep-th]].

\bibitem{Veneziano6}
P.~Di Vecchia, C.~Heissenberg, R.~Russo and G.~Veneziano,
  %``The gravitational eikonal: from particle, string and brane collisions to black-hole encounters,''
  Phys.\ Rept.\ {\bf 1083}, 1 (2024),
  [arXiv:2306.16488 [hep-th]].
%%%%%%%%%%%%%%%%%%%%%%%%%%%%%%%%%%%%%%%%%%%%%%%%%%%%%%%%%%%%%%%%%%%%%%%%%%%%%%%%%%%%%%

\bibitem{Veneziano7}
P.~Di Vecchia, C.~Heissenberg, R.~Russo and G.~Veneziano,
  %``Universality of ultra-relativistic gravitational scattering,''
  Phys.\ Lett.\ B {\bf 811}, 135924 (2020),
  [arXiv:2008.12743 [hep-th]].

\bibitem{Veneziano8}
P.~Di Vecchia, C.~Heissenberg, R.~Russo and G.~Veneziano,
  %``Radiation Reaction from Soft Theorems,''
  Phys.\ Lett.\ B {\bf 818}, 136379 (2021),
  [arXiv:2101.05772 [hep-th]].

\bibitem{Veneziano9}
P.~Di Vecchia, C.~Heissenberg, R.~Russo and G.~Veneziano,
  %``The eikonal operator at arbitrary velocities I: the soft-radiation limit,''
  JHEP {\bf 07}, 039 (2022),
  [arXiv:2204.02378 [hep-th]].

\bibitem{Veneziano10}
P.~Di Vecchia, C.~Heissenberg, R.~Russo and G.~Veneziano,
  %``Classical Gravitational Observables from the Eikonal Operator,''
  Phys.\ Lett.\ B {\bf 843}, 138049 (2023),
  [arXiv:2210.12118 [hep-th]].

\bibitem{SabVera}
A.~Sabio Vera, E.~Serna Campillo and M.~A.~Vazquez-Mozo,
  %``Graviton emission in Einstein-Hilbert gravity,''
  JHEP {\bf 03}, 005 (2012),
  [arXiv:1112.4494 [hep-th]].

\bibitem{SabVera1}
A.~Sabio Vera, E.~Serna Campillo and M.~A.~Vazquez-Mozo,
  %``Color-Kinematics Duality and the Regge Limit of Inelastic Amplitudes,''
  JHEP {\bf 04}, 086 (2013),
  [arXiv:1212.5103 [hep-th]].

\bibitem{SabVera2}
A.~Sabio Vera and M.~A.~Vazquez-Mozo,
  %``The Double Copy Structure of Soft Gravitons,''
  JHEP {\bf 03}, 070 (2015),
  [arXiv:1412.3699 [hep-th]].

\bibitem{SabVera3}
H.~Johansson, A.~Sabio Vera, E.~Serna Campillo and M.~\'A.~V\'azquez-Mozo,
  %``Color-Kinematics Duality in Multi-Regge Kinematics and Dimensional Reduction,''
  JHEP {\bf 10}, 215 (2013),
  [arXiv:1307.3106 [hep-th]].

%%%%%%%%%%%%%%%%%%%%%%%%%%%%%%%%%%%%%%%%%%%%%%%%%%%%%%%%%%%%%%%%%%%%%%%%%%%%%%%%%%%%%%
\bibitem{SabVera4}
A.~Sabio Vera,
  %``The double-logarithmic four-graviton Regge sector as a rank-two twisted period system,''
  JHEP {\bf 09}, 063 (2026),
  [arXiv:2604.05139 [hep-th]].

\bibitem{Venug}
H.~Raj and R.~Venugopalan,
  %``Universal features of 2 to N scattering in QCD and gravity from shockwave collisions,''
  Phys.\ Rev.\ D {\bf 109}, 044064 (2024),
  [arXiv:2311.03463 [hep-th]].

\bibitem{Venug1}
H.~Raj and R.~Venugopalan,
  %``Wee partons in QCD and gravity: double copy and universality,''
  EPJ Web Conf.\ {\bf 296}, 13013 (2024),
  [arXiv:2312.11652 [hep-ph]].

\bibitem{Venug2}
H.~Raj and R.~Venugopalan,
  %``QCD-Gravity double-copy in the Regge regime: shock wave propagators,''
  Phys.\ Rev.\ D {\bf 110}, 056010 (2024),
  [arXiv:2406.10483 [hep-th]].
%%%%%%%%%%%%%%%%%%%%%%%%%%%%%%%%%%%%%%%%%%%%%%%%%%%%%%%%%%%%%%%%%%%%%%%%%%%%%%%%%%%%%%

\bibitem{Venug3}
H.~Raj and R.~Venugopalan,
  %``Gravitational wave double copy of radiation from gluon shockwave collisions,''
  Phys.\ Lett.\ B {\bf 853}, 138669 (2024),
  [arXiv:2312.03507 [hep-th]].

\bibitem{Venug4}
H.~Raj and R.~Venugopalan,
  %``QCD-Gravity double copy in Regge asymptotics: from 2 to n amplitudes to radiation in shockwave collisions,''
  Acta Phys.\ Polon.\ B {\bf 56}, no.11, A1 (2025),
  [arXiv:2507.21252 [hep-th]].

\bibitem{Venug5}
A.~M.~Sta\'{s}to, H.~Raj and R.~Venugopalan,
  %``Squeezed-state radiation in shockwave scattering: QCD-Gravity double copy,''
  Acta Phys.\ Polon.\ B {\bf 57}, no.6, A3 (2026),
  [arXiv:2605.03038 [hep-th]].

\bibitem{Venug6}
I.~Blackstad, H.~Raj and R.~Venugopalan,
  %``Tidal effects on radiation in gravitational shockwave collisions,''
  [arXiv:2608.27585 [hep-th]].

%%%%%%%%%%%%%%%%%%%%%%%%%%%%%%%%%%%%%%%%%%%%%%%%%%%%%%%%%%%%%%%%%%%%%%%%%%%%%%%%%%%%%%%

\bibitem{Our13}
S.~Bondarenko and S.~Pozdnyakov,
  %``Effective Action and Classical Solutions,''
  Phys.\ Part.\ Nucl.\ Lett.\ {\bf 16}, no.5, 433 (2019).

%%%%%%%%%%%%%%%%%%%%%%%%%%%%%%%%%%%%%%%%%%%%%%%%%%%%%%%%%%%%%%%%%%%%%%%%%%%%%%%%%%%%%%%%%%%%%
% optical scalars, gravitational lensing, van Vleck determinant, Penrose limit

\bibitem{Sachs}
R.~K.~Sachs,
  %``Gravitational waves in general relativity. VI. The outgoing radiation condition,''
  Proc.\ Roy.\ Soc.\ Lond.\ A {\bf 264}, 309 (1961).

\bibitem{Ray}
A.~Raychaudhuri,
  %``Relativistic cosmology. I.,''
  Phys.\ Rev.\ {\bf 98}, 1123 (1955).

\bibitem{SEF}
P.~Schneider, J.~Ehlers and E.~E.~Falco,
  {\it Gravitational Lenses},
  Springer-Verlag, Berlin (1992).

\bibitem{SSE}
S.~Seitz, P.~Schneider and J.~Ehlers,
  %``Light propagation in arbitrary space-times and the gravitational lens approximation,''
  Class.\ Quant.\ Grav.\ {\bf 11}, 2345 (1994),
  [arXiv:astro-ph/9403056 [astro-ph]].

\bibitem{Perlick}
V.~Perlick,
  %``Gravitational lensing from a spacetime perspective,''
  Living Rev.\ Rel.\ {\bf 7}, 9 (2004).

\bibitem{VanVleck}
J.~H.~Van~Vleck,
  %``The correspondence principle in the statistical interpretation of quantum mechanics,''
  Proc.\ Nat.\ Acad.\ Sci.\ {\bf 14}, 178 (1928).

\bibitem{Morette}
C.~Morette,
  %``On the definition and approximation of Feynman's path integrals,''
  Phys.\ Rev.\ {\bf 81}, 848 (1951).

\bibitem{Visser}
M.~Visser,
  %``van Vleck determinants: Geodesic focusing and defocusing in Lorentzian space-times,''
  Phys.\ Rev.\ D {\bf 47}, 2395 (1993),
  [arXiv:hep-th/9303020 [hep-th]].

\bibitem{PPV}
E.~Poisson, A.~Pound and I.~Vega,
  %``The motion of point particles in curved spacetime,''
  Living Rev.\ Rel.\ {\bf 14}, 7 (2011),
  [arXiv:1102.0529 [gr-qc]].

\bibitem{Penrose}
R.~Penrose,
  {\it Any space-time has a plane wave as a limit},
  in {\it Differential Geometry and Relativity}, eds. M.~Cahen and M.~Flato,
  Reidel, Dordrecht (1976), p.~271.

\bibitem{Blau}
M.~Blau, M.~Borunda, M.~O'Loughlin and G.~Papadopoulos,
  %``Penrose limits and space-time singularities,''
  Class.\ Quant.\ Grav.\ {\bf 21}, L43 (2004),
  [arXiv:hep-th/0312029 [hep-th]].


\end{thebibliography}
\end{document}